\documentclass[final,a4paper,10pt]{article}
\usepackage{times}
\usepackage{amsmath}
\usepackage{amssymb}
\usepackage{amsfonts}
\usepackage{amsthm}
\usepackage{mathrsfs}
\usepackage{stmaryrd}
\usepackage[english]{babel}
\usepackage{color,graphicx}
\usepackage{a4wide}
\usepackage[all]{xy}
\usepackage{fancyvrb}

\theoremstyle{plain}
\newtheorem{theorem}{Theorem}[section]
\newtheorem{lemma}[theorem]{Lemma}
\newtheorem{corollary}[theorem]{Corollary}

\theoremstyle{definition}

\newtheorem{example}[theorem]{Example}
\newtheorem{example2}[theorem]{Example}

\theoremstyle{remark}

\numberwithin{equation}{section}

\theoremstyle{remark}

\newcommand{\comment} [1]{}

\def\ok#1{\mbox{\raisebox{0ex}[1ex][1ex]{$#1$}}}
\def \tuple#1{\langle #1 \rangle}

\newcommand{\sra}{{\shortrightarrow}}

\newcommand{\ACTL}{\ensuremath{\mathrm{ACTL}}}
\newcommand{\CTL}{\ensuremath{\mathrm{CTL}}}
\newcommand{\CTLS}{\ensuremath{\mathrm{CTL\!}^*}}
\newcommand{\ACTLS}{\ensuremath{\mathrm{ACTL\!}^*}}
\newcommand{\CTLX}{\ensuremath{\mathrm{CTL}\mbox{-}\mathrm{X}}}
\newcommand{\CTLSX}{\ensuremath{\mathrm{CTL\!}^*\mbox{-}\mathrm{X}}}

\newcommand{\hml}{\mathrm{HML}}

\newcommand{\ud}{\mbox{\raisebox{0ex}[1ex][1ex]{$\:\stackrel{{\scriptscriptstyle
\mathrm{def}}}{=}\:$}}}

\newcommand{\ra}{\rightarrow}
\newcommand{\Lra}{\Leftrightarrow}
\newcommand{\Ra}{\Rightarrow}
\newcommand{\La}{\Leftarrow}

\newcommand{\cK}{{\mathcal{K}}}

\newcommand{\cS}{{\mathcal{S}}}
\newcommand{\cX}{{\mathcal{X}}}

\newcommand{\cA}{{\mathcal{A}}}

\newcommand{\cM}{{\mathcal{M}}}

\newcommand{\cT}{{\mathcal{T}}}
\newcommand{\cP}{{\mathcal{P}}}

\newcommand{\fL}{\ensuremath{\mathcal{L}}}

\newcommand{\bZ}{\ensuremath{\mathbb{Z}}}

\newcommand{\sS}{\ensuremath{\mathscr{S}}}

\newcommand{\FM}{\ensuremath{F^{\scriptscriptstyle \cM}}}

\newcommand{\Op}{{\mathit{Op}}}

\newcommand{\beu}{{\bf EU}}
\newcommand{\bef}{{\bf EF}}

\newcommand{\pret}{\ensuremath{\widetilde{\pre}}}
\newcommand{\postt}{\ensuremath{\widetilde{\post}}}

\def\grasse#1{[\![#1]\!]}

\newcommand*{\gdca}{(\alpha ,C,A,\gamma)}

\DeclareMathOperator{\parent}{parent}
\DeclareMathOperator{\poset}{poset}
\DeclareMathOperator{\Absp}{Abs^{\mathrm{par}}}

\DeclareMathOperator{\dAbs}{dAbs}

\DeclareMathOperator{\dc}{dc}
\DeclareMathOperator{\id}{id}
\DeclareMathOperator{\sd}{\mathscr{S}_{\mathrm{dis}}}
\DeclareMathOperator{\pad}{pad}
\DeclareMathOperator{\D}{{\mathbb{D}}}

\DeclareMathOperator{\AP}{{\!\mathit{AP}\!}}
\DeclareMathOperator{\ari}{\sharp}
\DeclareMathOperator{\pre}{pre}

\DeclareMathOperator{\post}{post}
\DeclareMathOperator{\uco}{uco}

\DeclareMathOperator{\ucop}{uco^{par}}

\DeclareMathOperator{\lfp}{lfp}
\DeclareMathOperator{\gfp}{gfp}

\DeclareMathOperator{\img}{img}
\DeclareMathOperator{\Part}{Part}

\DeclareMathOperator{\Abs}{Abs}

\DeclareMathOperator{\Fun}{Fun}

\DeclareMathOperator{\pr}{par}

\DeclareMathOperator{\fin}{\mathit{fin}}
\DeclareMathOperator{\ptsplit}{PTsplit}
\DeclareMathOperator{\gvsplit}{GVsplit}

\DeclareMathOperator{\splitters}{Splitters}
\DeclareMathOperator{\gvrefiners}{GVrefiners}
\DeclareMathOperator{\ptrefiners}{PTrefiners}
\DeclareMathOperator{\ptblockrefiners}{PTblockrefiners}
\DeclareMathOperator{\PTrefiners}{PTrefiners}
\DeclareMathOperator{\refiners}{Refiners}
\DeclareMathOperator{\subrefiners}{subRefiners}
\DeclareMathOperator{\blockrefiners}{BlockRefiners}

\DeclareMathOperator{\Refiners}{Refiners}

\DeclareMathOperator{\refine}{refine}

\DeclareMathOperator{\PT}{PT}
\DeclareMathOperator{\GV}{GV}

\DeclareMathOperator{\GPT}{GPT}

\DeclareMathOperator{\IGPT}{IGPT}
\DeclareMathOperator{\CPT}{CPT}

\DeclareMathOperator{\pos}{pos}

\DefineVerbatimEnvironment%
{code}{Verbatim}{commandchars=\\\[\],
codes={\catcode`$=3\catcode`^=7\catcode`_=8},
numbers=left,frame=lines}

\newcommand{\kforall}{{\bf{for all}}}

\newcommand{\kdo}{{\bf{do}}}
\newcommand{\kcontinue}{{\bf{continue}}}
\newcommand{\kin}{{\bf in}}

\newcommand{\kif}{{\bf{if}}}
\newcommand{\kthen}{{\bf then}}
\newcommand{\kelse}{{\bf else}}

\newcommand{\knew}{{\bf{new}}}
\newcommand{\kdelete}{{\bf{delete}}}

\begin{document}

\title{\Large \bf Generalizing the Paige-Tarjan Algorithm\\
by Abstract Interpretation}

\author{\normalsize {\sc Francesco Ranzato} ~~~ {\sc Francesco Tapparo}\\
\normalsize Dipartimento di Matematica Pura ed Applicata, Universit\`a
di Padova\\
\normalsize Via Belzoni 7, 35131 Padova, Italy\\
\normalsize \texttt{ranzato$@$math.unipd.it} ~~~~ \texttt{tapparo$@$math.unipd.it}
}

\date{}
\pagestyle{plain}

\maketitle

\begin{abstract}
The Paige and Tarjan algorithm ($\PT$) for computing the coarsest
refinement of a state partition which is a bisimulation on some Kripke
structure is well known.  It is also well known in model
checking that bisimulation is equivalent to strong preservation of $\CTL$
or, equivalently, of Hennessy-Milner logic.  Drawing on these observations,
we analyze the basic steps of the $\PT$ algorithm from an abstract
interpretation perspective, which allows us to reason on strong
preservation in the context of generic inductively defined (temporal)
languages and of possibly non-partitioning abstract models specified by abstract
interpretation. This leads us to design a generalized Paige-Tarjan
algorithm, called $\GPT$, for computing the minimal refinement of an
abstract interpretation-based model that strongly preserves some given
language.  It turns out that $\PT$ is a straight instance of $\GPT$
on the domain of state partitions for the case of strong preservation
of Hennessy-Milner logic. We provide a number of examples showing that
$\GPT$ is of general use. We first show how a well-known efficient algorithm
for computing stuttering equivalence can be viewed as a
simple instance of $\GPT$.  We then instantiate $\GPT$ in order to
design a new efficient algorithm for computing simulation equivalence
that is competitive with the  best
available algorithms. Finally, we show how $\GPT$ allows to compute new
strongly  preserving abstract models by providing an efficient algorithm
that computes the coarsest refinement of a given partition that
strongly preserves the language generated by the reachability operator.\\[10pt]
\end{abstract}

\section{Introduction}\label{intro}

\paragraph*{Motivations.} 
The Paige and Tarjan~\cite{pt87} algorithm~---~in the paper denoted by
$\PT$~---~for efficiently computing the coarsest refinement of a given
partition which is
\emph{stable} for a given state transition relation is well known. Its importance stems
from the fact that $\PT$ actually computes
\emph{bisimulation equivalence}, because a partition $P$ of a state space
$\Sigma$ is stable for a transition relation $\sra$ on $\Sigma$
if and only if
$P$ is a bisimulation equivalence on the transition system $\tuple{\Sigma,\sra}$. In
particular, $\PT$ is widely used in model checking for reducing the
state space of a Kripke structure $\cK$
because the quotient of $\cK$ w.r.t.\ 
bisimulation equivalence \emph{strongly preserves} temporal
languages like $\CTLS$, $\CTL$ and the whole $\mu$-calculus
\cite{bcg88,cgp99}. This means that logical specifications can be
checked on the abstract quotient model of $\cK$ with no loss of precision. 
Paige and Tarjan first present the basic $O(|\sra|
|\Sigma|)$-time $\PT$ algorithm  and 
then exploit a computational logarithmic improvement in order to
design a $O(|\sra|
\log |\Sigma|)$-time algorithm, which is usually referred to as 
Paige-Tarjan algorithm. It is important to remark that the logarithmic
Paige-Tarjan algorithm is derived as an algorithmic refinement of
$\PT$ that does not affect the correctness of the procedure which is
instead proved for the basic $\PT$ algorithm.  As shown in \cite{rt06}, it
turns out that state partitions can be viewed as \emph{domains in abstract
interpretation} and strong preservation
can be cast as \emph{completeness in abstract interpretation}. Thus, our
first aim was to make use of an ``abstract interpretation eye'' to
understand why $\PT$ is a correct procedure for computing strongly
preserving partitions. 
\paragraph*{The PT Algorithm.} 
Let us recall how
$\PT$ works. Let 
$\pre_\sra(X)=\{s\in \Sigma~|~\exists x\in X.\, s
\ok{\sra} x\}$ 
denote the usual predecessor transformer on $\wp(\Sigma)$.
A partition $P\in \Part(\Sigma)$ is 
$\PT$ stable when for any block $B\in P$, if $B'\in P$ then 
either $B\subseteq
\pre_\sra(B')$ or $B\cap \pre_\sra(B')=\varnothing$.
For a given subset $S\subseteq \Sigma$, $\ptsplit(S,P)$ denotes the partition
obtained from $P$ by replacing each block $B\in P$ with the 
blocks $B\cap\pre_\sra(S)$ and 
$B\smallsetminus\pre_\sra(S)$, where we
also allow no splitting, that is, $\ptsplit(S,P)=P$. 
When $P\neq \ptsplit(S,P)$ the subset $S$ is called
a \emph{splitter} for $P$. $\splitters(P)$ denotes the set of splitters of
$P$, while $\ptrefiners(P)\ud  \{ S\in \splitters(P)~|~ \exists \{B_i
\}\subseteq P.\: S=\cup_i B_i \}$. 
Then, the $\PT$ algorithm goes as follows.

\begin{center}
{\small
$
\begin{array}{|l|}
\hline \\[-9pt]
~\mbox{{\bf input}}\!: \text{partition~} P\in \Part(\Sigma);~~ \\
~\mbox{{\bf while~}} (P \text{~is~not~$\PT$~stable}) ~\mbox{{\bf do}}~~\\
~~~~~~\mbox{{\bf choose~}} S \in \ptrefiners(P);~~ \\
~~~~~~P:=\ptsplit(S,P);\\
~\mbox{{\bf endwhile}}\\[-3pt]
~\mbox{{\bf output}}\!:~ P;
~~~~~~~~~~~~~~~~~~~~~~~~~~~\framebox{$\PT$}\mbox{\hspace*{-5pt}}\\[-0.35pt]
\hline
\end{array}
$
}
\end{center}

\noindent The time complexity of $\PT$ is $O(|\sra|
|\Sigma|)$ because the number of while loops is bounded by $|\Sigma|$
and, by storing $\pre_\sra(\{s\})$ for each $s\in \Sigma$, finding a
$\PT$ refiner and performing the splitting takes $O(|\sra|)$ time. 

\paragraph*{An Abstract Interpretation Perspective of PT.}
This work originated from a number of observations on the above $\PT$ algorithm. 
Firstly, we may view the output $\PT(P)$ as the
coarsest refinement of a partition $P$ that strongly preserves
$\CTL$. For partitions of the state space $\Sigma$, namely
standard abstract models in model checking, it is
known that strong preservation of $\CTL$ is equivalent to strong
preservation of (finitary) Hennessy-Milner logic $\hml$ \cite{hm85}, i.e., the
language: $$\varphi ::= p ~|~ \varphi_1
\wedge \varphi_2 ~|~ \neg \varphi ~|~
\mathrm{EX}\varphi$$ 
The interpretation of $\hml$ is standard: $p$ ranges
over atomic propositions in $\AP$ where $\{\grasse{p}\subseteq
\Sigma~|~p\in \AP\}=P$ and the semantic
interpretation of the existential next operator $\mathrm{EX}$ is 
$\pre_\sra : \wp(\Sigma)\ra \wp(\Sigma)$. We observe that $\PT(P)$ indeed
computes the coarsest partition $P_{\scriptscriptstyle \hml}$ that
refines $P$ and strongly preserves $\hml$. Moreover, the partition
$P_{\scriptscriptstyle \hml}$ 
corresponds to
the state equivalence $\equiv_{\scriptscriptstyle \hml}$ induced by
the semantics of $\hml$: $s\equiv_{\scriptscriptstyle \hml}
s'$ iff $\forall \varphi \in
\hml.\, s\in
\grasse{\varphi} \Lra s' \in \grasse{\varphi}$. We also observe
that $P_{\scriptscriptstyle
\hml}$ is an
abstraction on the domain
$\Part(\Sigma)$ of
partitions of $\Sigma$
of the standard state semantics of $\hml$. 
Thus, our starting point was that $\PT$ can be viewed as an
algorithm for computing the most abstract object 
on a particular domain, i.e.\ $\Part(\Sigma)$, that strongly
preserves a particular language, i.e.\
$\hml$. We make this view precise within Cousot and
Cousot's abstract interpretation framework \cite{CC77,CC79}.  \\
\indent
Previous work \cite{rt06} introduced an abstract interpretation-based
framework for reasoning on strong preservation of abstract models
w.r.t.\ generic inductively defined languages. We showed that the
lattice $\Part(\Sigma)$ of partitions of the state space $\Sigma$ can
be viewed as an abstraction, through some abstraction and
concretization maps $\alpha$ and $\gamma$, of the lattice $\Abs(\wp(\Sigma))$ of
abstract interpretations of $\wp(\Sigma)$. Thus, a partition $P\in
\Part(\Sigma)$ is here viewed as a particular abstract domain
$\gamma(P)\in \Abs(\wp(\Sigma))$.  This leads to a precise
correspondence between \emph{forward complete} abstract
interpretations and strongly preserving abstract models. Let us recall
that completeness in abstract interpretation \cite{CC77,CC79,grs00}
encodes an ideal situation where no loss of precision occurs by
approximating concrete computations on abstract domains.  The
problem of minimally refining an abstract model in order to get strong
preservation of some language $\fL$ can be cast as the problem of
making an abstract interpretation $\cA$ forward complete for the
semantic operators of $\fL$ through a minimal refinement of the
abstract domain of $\cA$. It turns out that this latter completeness
problem always admits a fixpoint solution.  Hence, in our abstract
interpretation framework, it turns out that for any $P\in
\Part(\Sigma)$, the output $\PT(P)$ is the partition abstraction
in
$\Part(\Sigma)$ through $\alpha$ of the minimal refinement of the
abstract domain $\gamma(P)\in
\Abs(\wp(\Sigma))$ that is complete for the set $\Op_{\scriptscriptstyle \hml}$ of semantic
operators of the language $\hml$, where $\Op_{\scriptscriptstyle \hml}
=\{\cap, \complement, \pre_\sra\}$ therefore includes intersection, complementation
and precedessor operators. In particular, a partition
$P$ is $\PT$ stable iff the abstract domain $\gamma(P)$ is complete for the operators in
$\Op_{\scriptscriptstyle \hml}$.  Also, the following observation is
crucial in our approach. The splitting operation $\ptsplit(S,P)$ can
be viewed as the \emph{best correct approximation} on $\Part(\Sigma)$
of a refinement operation $\refine_{\mathit{op}}(S,\cdot)$ of abstract domains:
given an operator $\mathit{op}$, $\refine_{\mathit{op}}(S,A)$ refines an
abstract domain $A$ through a ``$\mathit{op}$-refiner'' $S \in A$ to the most
abstract domain that contains both $A$ and the image $\mathit{op}(S)$.  In particular, $P$
results to be $\PT$ stable iff the abstract domain $\gamma(P)$ cannot
be refined w.r.t.\
the function $\pre_\sra$. Thus, if
$\ok{\refine_{\mathit{op}}^{\scriptscriptstyle \Part}}$  denotes the
best correct approximation in $\Part(\Sigma)$ of $\refine_{\mathit{op}}$ then the
$\PT$ algorithm can be reformulated as follows.
\begin{center}
{\small
$
\begin{array}{|l|}
\hline \\[-9pt]
~\mbox{{\bf input}}\!: \text{partition~} P\in \Part(\Sigma);~~ \\
~\mbox{{\bf while~}} \text{~the set of $\pre_\sra$-refiners of $P$ $\neq
\varnothing$ } ~\mbox{{\bf do}}~\\
~~~~~~\mbox{{\bf choose~}} \text{some $\pre_\sra$-refiner $S\in \gamma(P)$};~ \\
~~~~~~P :=\refine^{\scriptscriptstyle \Part}_{\scriptscriptstyle \pre_\sra}(S,P);~\\
~\mbox{{\bf endwhile}}\\[-0.5pt]
~\mbox{{\bf output}}\!:~ P; \\
\hline
\end{array}
$
}
\end{center}

\paragraph*{Main Results.}
This abstract interpretation-based view of $\PT$ leads us to
generalize $\PT$ to: 
\begin{itemize}
\item[{\rm (1)}] a generic domain $\cA$ of abstract models that
generalizes the role played in $\PT$ by the domain of state partitions $\Part(\Sigma)$; 
\item[{\rm (2)}] a
generic set $\Op$ of operators on $\wp(\Sigma)$ that provides the semantics
of some language $\fL_{\Op}$ and generalizes the role played in $\PT$ by the set
$\Op_{\scriptscriptstyle \hml}$ of operators of $\hml$. 
\end{itemize}

We design a
generalized Paige-Tarjan refinement algorithm, called $\GPT$, that,
for any abstract model $A\in \cA$, computes the most
abstract refinement of $A$ in $\cA$ which is strongly preserving for
the language $\fL_{\Op}$. The correctness of $\GPT$ is guaranteed by some
completeness conditions on $\cA$ and $\Op$.  We provide a number of
applications showing that $\GPT$ is an algorithmic scheme of general
use.

We first show how $\GPT$ can be instantiated in order to get the
well-known Groote-Vaandrager algorithm~\cite{gv90} that computes
divergence blind stuttering equivalence in $O(|\sra||\Sigma|)$-time
(this is the best known time bound). Divergence blind stuttering
equivalence is a behavioural equivalence used in process algebra to
take into account invisible events \cite{bcg88,dnv95}.  Let us recall
that the Groote-Vaandrager algorithm can be also used for computing
branching bisimulation equivalence, which is the state equivalence
induced by $\CTLSX$
\cite{bcg88,dnv95,gv90}. 
The Groote-Vaandrager algorithm corresponds to an instance of $\GPT$ where
the set of operators is $\Op=\{\cap,\complement,
\mathbf{EU}\}$~--~$\mathbf{EU}$ denotes the standard semantic
interpretation of the existential until~--~and the abstract domain
$\cA$ is the lattice of partitions $\Part(\Sigma)$. 

We then show how $\GPT$ allows to design a new simple and efficient
algorithm for computing simulation equivalence. This algorithm is
obtained as a consequence of the fact that simulation equivalence
corresponds to strong preservation of the language $$\varphi ::= p ~|~
\varphi_1 \wedge \varphi_2 ~|~
\mathrm{EX}\varphi.$$ 
Therefore,  in this instance of $\GPT$ 
the set of operators is $\Op=\{\cap,\pre_\sra\}$ and the abstract domain
$\cA$ is the lattice of disjunctive (i.e.\ precise for least upper
bounds \cite{CC79}) abstract domains of $\wp(\Sigma)$. 
It turns out that this algorithm can be
implemented with
space and time complexities that are competitive with
those of the best available algorithms for simulation equivalence. 

Finally, we demonstrate how $\GPT$ can solve 
novel strong preservation problems by considering strong
preservation w.r.t.\ the language inductively generated by
propositional logic and the reachability operator $\bef$. Here, we
obtain a partition refinement algorithm, namely the abstract domain
$\cA$ is the lattice of partitions $\Part(\Sigma)$, while the set of
operators is $\Op=\{\cap,\complement,
\mathbf{EF}\}$. 
We describe an implementation for this instance of $\GPT$ 
that leads to a 
$O(|\sra||\Sigma|)$-time algorithm that was also experimentally evaluated.

\section{Background}

\subsection{Notation and Preliminaries} 

\paragraph{Notations.} 
Let $X$ be any set.
$\Fun(X)$ denotes the set of functions $f:X^n\ra X$, for any $n
=\ari(f) \geq
0$, called arity of $f$.  Following a standard convention, when $n=0$,
$f$ is meant to be a specific object of $X$. 
If $f:X\ra Y$ then the
image of $f$ is also denoted by $\img(f)=\{f(x)\in Y~|~ x\in X\}$. 
When writing a set $S$ of
subsets of a given set, like a partition, $S$ is often written in a
compact form like $\{1, 12, 13\}$ or $\{[1], [12], [13]\}$ that stands
for $\{\{1\}, \{1,2\}, \{1,3\}\}$.
The complement operator for the universe set $X$ is $\complement
:\wp(X)\ra \wp(X)$, where $\complement (S)=X\smallsetminus S$.

\paragraph{Orders.} 
Let $\tuple{P,\leq}$ be a poset. Posets are often denoted by $P_\leq$.
We use the symbol ($\sqsubset$) $\sqsubseteq$ to denote (strict) pointwise ordering between
functions: If $X$ is any set and $f,g:X \ra P$ then $f\sqsubseteq g$
if for all $x\in X$, $f(x)\leq g(x)$.  A mapping $f: P\ra Q$ on posets
is continuous when $f$ preserves least upper bounds (lub's) of
countable chains in $P$, while, dually, it is co-continuous when $f$
preserves greatest lower bounds (glb's) of countable chains in $P$. A
complete lattice $C_\leq$ is also denoted by
$\tuple{C,\leq,\vee,\wedge,\top,\bot}$ where $\vee$, $\wedge$, $\top$
and $\bot$ denote, respectively, lub, glb, greatest element and least
element in $C$.  
A function $f:C\ra D$ between complete lattices is additive
(co-additive) when $f$
preserves least upper (greatest lower) bounds.   
We denote by
$\lfp(f)$ and $\gfp (f)$, respectively, the least and greatest
fixpoint, when they exist, of an operator $f$ on a poset.

\paragraph{Partitions.} A partition $P$ of a set $\Sigma$ is a set of
nonempty subsets of $\Sigma$, 
called
blocks, that are pairwise disjoint and whose union gives $\Sigma$.
$\Part(\Sigma)$ denotes the set of partitions of $\Sigma$.  $\Part(\Sigma)$ is
endowed with the following standard partial order $\preceq$: $P_1
\preceq P_2$, i.e.\ $P_2$ is coarser than $P_1$ (or $P_1$ refines
$P_2$) iff $\forall B\in P_1. \exists B' \in P_2 .\: B \subseteq
B'$. It is well known that $\tuple{\Part(\Sigma),\preceq,
\curlywedge, \curlyvee, \{\Sigma\}, \{\{s\}\}_{s\in \Sigma}}$ is a complete
lattice, where $P_1 \curlywedge P_2 = \{B_1 \cap B_2 ~|~ B_1\in P_1,\: B_2 \in
P_2,\: B_1 \cap B_2 \neq \varnothing\}$.

\paragraph{Kripke Structures.}
A transition system $\cT=(\Sigma ,\sra)$ consists of a (possibly
infinite) set $\Sigma$ of states and a transition relation $\sra 
\subseteq \Sigma \times
\Sigma$. 
As usual \cite{cgp99}, we assume that the relation $\sra$ is total,
i.e., for any $s\in \Sigma $ there exists some $t\in \Sigma $ such
that $s\sra t$, so that any maximal path in $\cT$ is necessarily
infinite.  The pre/post
transformers on $\wp(\Sigma)$ are defined as usual:

\medskip
$
\begin{array}{lll}
\text{--}\!\!& \pre_\sra \ud \lambda Y. \{ a\in \Sigma ~|~\exists b\in Y.\; a \sra
b\} \\[5pt]
\text{--}\!\!& \pret_\sra \ud \complement \circ \pre_\sra \circ \complement = 
\lambda Y. \{ a\in \Sigma  ~|~\forall b\in \Sigma . (a \sra
b \Rightarrow  b\in Y)\} \\[5pt] 
\text{--}\!\!& \post_\sra \ud \lambda Y. \{ b\in \Sigma ~|~\exists a\in Y.\; a \sra
b\} \\[5pt]
\text{--}\!\!& \postt_\sra \ud \complement \circ \post_\sra \circ \complement = 
\lambda Y. \{ b\in \Sigma ~|~\forall a\in \Sigma . (a \sra
b \Rightarrow a\in Y)\} 
\end{array}
$

\medskip
\noindent
Let us remark that $\pre_\sra$ and $\post_\sra$ are additive operators on
$\wp(\Sigma)_\subseteq$ while $\pret_\sra$ and
$\postt_\sra$ are co-additive. When clear from the context, subscripts
in pre/post transformers are sometimes omitted.

Given a set $\mathit{AP}$ of atomic propositions
(of some language), a Kripke structure $\cK= (\Sigma ,\sra,\ell)$ over
$\mathit{AP}$ consists of a transition system $(\Sigma ,\sra)$
together with a state labeling function $\ell:\Sigma \ra
\wp(\mathit{AP})$. We use the following notation: for any $s\in
\Sigma$, $[s]_\ell \ud
\{s'\in \Sigma~|~ \ell(s)=\ell(s')\}$, while $P_\ell \ud \{[s]_\ell~|~
s\in \Sigma\}\in \Part(\Sigma)$ denotes the state partition that is induced by
$\ell$.  

The notation $s \ok{\models^\cK} \varphi$ means
that a state $s\in \Sigma$ satisfies in $\cK$ a state formula $\varphi$ of
some language $\fL$, where the specific definition of the satisfaction
relation $\ok{\models^\cK}$ depends on the language $\fL$
(interpretations of standard logical/temporal operators like next,
until, globally, etc.\ can be found
in \cite{cgp99}).

\subsection{Abstract Interpretation and Completeness}\label{aic}

\subsubsection{Abstract Domains}\label{absdom}

In standard Cousot and Cousot's abstract interpretation, abstract
domains can be equivalently specified either by Galois connections,
i.e.\ adjunctions, or by upper closure operators
(uco's)~\cite{CC77,CC79}.  Let us recall these standard notions.

\paragraph{Galois Connections and Insertions.}
If $A$ and $C$ are posets and $\alpha :C\ra A$ and $\gamma :A\ra C$
are monotone functions such that $\forall c\in C.\: c \leq_C \gamma
(\alpha (c))$ and $\alpha(\gamma (a)) \leq_A a$ then the quadruple
$\gdca$ is called a Galois connection (GC for short) between $C$ and
$A$.  If in addition $\alpha\circ \gamma =\lambda x.x$ then $\gdca$ is
a Galois insertion (GI for short) of $A$ in $C$.  In a GI, $\gamma$ is
1-1 and $\alpha$ is onto.  Let us also recall that the notion of GC is
equivalent to that of adjunction: if $\alpha :C\ra A$ and $\gamma
:A\ra C$ then $\gdca$ is a GC iff $\forall c\in C.\forall a\in A.\;
\alpha (c)\leq_{A} a \Lra c\leq_C \gamma (a)$.  The map $\alpha$
($\gamma$) is called the left- (right-) adjoint to $\gamma$
($\alpha$).  It turns out that one adjoint map $\alpha$/$\gamma$
uniquely determines the other adjoint map $\gamma$/$\alpha$ as
follows. On the one hand, a map $\alpha:C\ra A$ admits a necessarily
unique right-adjoint map $\gamma:A\ra C$ iff $\alpha$ preserves
arbitrary lub's; in this case, we have that $\gamma \ud \lambda a.
\vee_C \{ c\in C~|~ \alpha(c) \leq_A a\}$. 
On the other hand, a map $\gamma:A\ra C$ admits a necessarily unique
left-adjoint map $\alpha:C\ra A$ iff $\gamma$ preserves arbitrary
glb's; in this case, $\alpha \ud \lambda c. \wedge_A \{ a\in A~|~ c
\leq_C \gamma(a)\}$.  In particular, in any GC $\gdca$ between
complete lattices it turns out that $\alpha$ is additive and $\gamma$
is co-additive.

We assume the standard abstract interpretation framework, where
concrete and abstract domains, $C$ and $A$, are complete lattices
related by abstraction and concretization maps $\alpha$ and $\gamma$
forming a GC $\gdca$. $A$ is called an abstraction of $C$ and $C$ a
concretization of $A$.  The ordering relations on concrete and
abstract domains describe the relative precision of domain values:
$x\leq y$ means that $y$ is an approximation of $x$ or, equivalently,
$x$ is more precise than $y$.  Galois connections relate the
concrete and abstract notions of relative precision: an abstract value
$a\in A$ approximates a concrete value $c\in C$ when $\alpha(c) \leq_A
a$, or, equivalently (by adjunction), $c\leq_C \gamma(a)$. As a key
consequence of requiring a Galois connection, it turns out that
$\alpha(c)$ is the best possible approximation in $A$ of $c$, that is
$\alpha(c) = \wedge
\{a\in A~|~ c  \leq_C \gamma(a)\}$ holds. 
If $\gdca$ is a GI then each value of the abstract domain $A$ is
useful in representing $C$, because all the values in $A$ represent
distinct members of $C$, being $\gamma$ 1-1.  Any GC can be lifted to
a GI by identifying in an equivalence class those values of the
abstract domain with the same concretization.  $\Abs(C)$ denotes the
set of abstract domains of $C$ and we write $A\in \Abs(C)$ to mean
that the abstract domain $A$ is related to $C$ through a GI
$(\alpha,C,A,\gamma)$. 

An abstract domain $A\in \Abs(C)$ is disjunctive when the
corresponding concretization map $\gamma$ is additive or,
equivalently, when the image $\gamma(A)\subseteq C$ is closed under
arbitrary lub's of $C$. We denote by $\dAbs(C)$ the subset of disjunctive abstract
domains.

\paragraph{Closure Operators.}
An (upper) closure operator, or simply a closure, on a poset $P_\leq$
is an operator $\mu: P\rightarrow P$ that is monotone, idempotent and
extensive, i.e., $\forall x\in P.\; x\leq
\mu (x)$.  Dually, lower closure operators 
are monotone, idempotent, and restrictive, i.e., $\forall x\in P.\;
\mu (x) \leq x$.  
$\uco(P)$ denotes the set of closure operators on $P$.
Let $\tuple{C, \leq ,\vee ,\wedge ,\top ,\bot}$ be a complete lattice.
A closure $\mu
\in\uco(C)$ is uniquely determined by its image $\img(\mu)$, which 
coincides with its set of fixpoints, as follows: $\mu=
\lambda y.\wedge \{ x\in \img(\mu)~|~ y\leq x\}$. Also, 
$X\subseteq C$ is the image of some closure operator $\mu_X$ on $C$
iff $X$ is a Moore-family of $C$, i.e., $X=\cM (X)\ud \{\wedge S~|~
S\subseteq X\}$~---~where $\wedge \varnothing=\top \in \cM (X)$. In
other terms, $X$ is a Moore-family of $C$ (or Moore-closed) when $X$ is meet-closed.  In
this case, $\mu_X=\lambda y.\wedge \{ x\in X~|~ y\leq x\}$ is the
corresponding closure operator on $C$.  For any $X\subseteq C$, $\cM
(X)$ is called the Moore-closure of $X$ in $C$, i.e., $\cM (X)$ is the
least (w.r.t.\ set inclusion) subset of $C$ which contains $X$ and is
a Moore-family of $C$. Moreover, it turns out that for any $\mu \in
\uco(C)$ and any Moore-family $X\subseteq C$, $\mu_{\img(\mu)} = \mu$
and $\img(\mu_X)=X$. Thus, closure operators on $C$ are in bijection
with Moore-families of $C$. This allows us to consider a closure
operator $\mu\in \uco(C)$ both as a function $\mu:C\ra C$ and as a
Moore-family $\img(\mu)\subseteq C$.  This is particularly useful and
does not give rise to ambiguity since one can distinguish the use of a
closure $\mu$ as function or set according to the context.


%
If $C$ is a complete lattice then $\uco(C)$ endowed with the pointwise ordering
$\sqsubseteq$ is a
complete lattice denoted by
$\tuple{\uco(C),\sqsubseteq,\sqcup,\sqcap,\lambda x.\top,\lambda x.x}$,
where for every $\mu,\eta \in\uco(C)$, $\{
\mu_i \}_{i\in I} \subseteq\uco(C)$ and $x\in C$:
\begin{itemize}
\item[--] $\mu \sqsubseteq \eta$ iff $\forall y\in C.\; \mu (y) \leq
 \eta(y)$ iff  $\img(\eta)
\subseteq \img(\mu)$;
\item[--] $(\sqcap_{i\in I} \mu_i)(x) = \wedge_{i\in I} \mu_i (x)$;
\item[--] $x \in \sqcup_{i\in I} \mu_i \:\Lra\:
\forall i\in I.\; x\in \img(\mu_i)$;
\item[--] $\lambda x.\top$ is the greatest element, whereas $\lambda x.x$ is
the least element.
\end{itemize}
Thus, the glb in $\uco(C)$ is defined pointwise, while the
lub of a set of closures $\{
\mu_i \}_{i\in I} \subseteq\uco(C)$ is the closure whose image 
is given by the set-intersection $\cap_{i \in I}
\mu_i$.

A closure $\mu\in \uco(C)$ is disjunctive when
$\mu$ preserves arbitrary lub's or, equivalently, when $\img(\mu)$ is
join-closed, that is $\{ \vee X ~|~ X
\subseteq \img(\mu)\}=\img(\mu)$.  Hence, a subset $X\subseteq C$ is the image
of a disjunctive closure on $C$ iff $X$ is both meet- and join-closed.  If
$C$ is completely distributive~---~this is the case, for example, of
a lattice $\tuple{\wp(\Sigma),\subseteq}$ for some set
$\Sigma$~---~then the greatest (w.r.t.\ $\sqsubseteq$) disjunctive
closure $\D(S)$ that contains a set $S\subseteq C$ is obtained by
closing $S$ under meets and joins, namely $\D(S)\ud \{\vee X~|~ X
\subseteq \cM(S)\}$.

\paragraph{Closures are Equivalent to Galois Insertions.}
It is well known since \cite{CC79} that abstract domains can be
equivalently specified either as Galois insertions or as closures.
These two approaches are completely equivalent. On the one hand, if
$\mu\in\uco(C)$ and $A$ is a complete lattice which is isomorphic to
$\img(\mu)$, where $\iota : \img(\mu)\ra A$ and $\iota^{-1}:
A\ra\img(\mu)$ provide the isomorphism, then
$(\iota\circ\mu,C,A,\iota^{-1} )$ is a GI. On the other hand, if
$\gdca$ is a GI then $\mu_A \ud \gamma \circ \alpha\in\uco(C)$ is the
closure associated with $A$ such that $\tuple{\img(\mu_A),\leq_C}$ is
a complete lattice which is isomorphic to
$\tuple{A,\leq_A}$. Furthermore, these two constructions are inverse
of each other.  Let us also remark that an abstract domain $A$ is
disjunctive iff the uco $\mu_A$ is disjunctive. Given an abstract domain $A$
specified by a GI $\gdca$, its associated closure $\gamma\circ\alpha$
on $C$ can be thought of as the ``logical meaning'' of $A$ in $C$,
since this is shared by any other abstract representation for the
objects of $A$.  Thus, the closure operator approach is particularly
convenient when reasoning about properties of abstract domains
independently from the representation of their objects.

\paragraph{The Lattice of Abstract Domains.}
Abstract domains specified by GIs can be pre-ordered w.r.t.\ precision
as follows: if $A_1,A_2 \in \Abs(C)$ then $A_1$ is more precise (or
concrete) than $A_2$ (or $A_2$ is an abstraction of $A_1$)
when $\mu_{A_1} \sqsubseteq \mu_{A_2}$.  The
pointwise ordering $\sqsubseteq$ between uco's corresponds therefore
to the standard ordering used to compare abstract domains with respect
to their precision. Also, $A_1$ and $A_2$ are equivalent, denoted by
$A_1 \simeq A_2$, when their associated closures coincide, i.e.\
$\mu_{A_1}=\mu_{A_2}$.  Hence, the quotient $\Abs(C)_{/\simeq}$ gives
rise to a poset that, by a slight abuse of notation, is simply denoted
by \mbox{$\tuple{\Abs(C),\sqsubseteq}$}. Thus, when we write $A\in\Abs(C)$ we
mean that $A$ is any representative of an equivalence class in
$\Abs(C)_{/\simeq}$ and is specified by a Galois insertion $\gdca$.
It turns out that $\tuple{\Abs(C),\sqsubseteq}$ is a complete lattice,
called the lattice of abstract domains of $C$
\cite{CC77,CC79},  because it is
isomorphic to the complete lattice $\tuple{\uco(C),\sqsubseteq}$.
Lub's and glb's in $\Abs(C)$ have therefore the following reading as
operators on domains.  Let $\{ A_i\}_{i\in I}\subseteq\Abs(C)$:
(i)~$\sqcup_{i\in I}A_i $ is the most concrete among the domains which
are abstractions of all the $A_i$'s; (ii)~$\sqcap_{i\in I} A_i $ is
the most abstract among the domains which are more concrete than every
$A_i$~---~this latter domain is also known as reduced product
\cite{CC79} of all the $A_i$'s.

\subsubsection{Completeness in Abstract Interpretation}\label{secco}

\paragraph{Correct Abstract Interpretations.}
Let $C$ be a concrete domain, $f:C\ra C$ be a concrete semantic function\footnote{For
simplicity of notation we consider here unary functions since the extension
to generic $n$-ary functions is straightforward.}  
and  $f^\sharp:A \ra A$ be a corresponding abstract
function on an abstract domain $A\in \Abs(C)$ specified by a GI 
$(\alpha,C,A,\gamma)$.   Then,
$\tuple{A,f^\sharp}$ is a sound (or correct) abstract interpretation when $\ok{\alpha
\circ f \sqsubseteq f^\sharp\circ \alpha}$ holds. The abstract
function $\ok{f^\sharp}$ is called a correct approximation on $A$ of
$f$.  This means that a concrete computation $f(c)$ can be correctly
approximated in $A$ by $\ok{f^\sharp (\alpha (c))}$, namely
$\alpha(f(c))
\leq_A \ok{f^\sharp (\alpha (c))}$. An abstract function 
$\ok{f_1^\sharp}:A \ra A$ is more precise than 
$\ok{f_2^\sharp}:A \ra A$ when 
$\ok{f_1^\sharp} \sqsubseteq \ok{f_2^\sharp}$. 
Since  $\ok{\alpha
\circ f \sqsubseteq f^\sharp\circ \alpha}$ holds iff  $\ok{\alpha
\circ f \circ \gamma \sqsubseteq f^\sharp}$ holds, 
the abstract function
$\ok{f^A \ud \alpha \circ f \circ \gamma: A\rightarrow A}$ is called the best correct
approximation of $f$ in $A$.  

\paragraph{Complete Abstract Interpretations.}
Completeness in abstract interpretation
corresponds to requiring that, in addition to soundness, no loss of
precision occurs when $f(c)$ is approximated in $A$  
by $\ok{f^\sharp(\alpha(c))}$. Thus, completeness of $\ok{f^\sharp}$
for $f$ is encoded by the equation $\ok{\alpha \circ f = f^\sharp \circ
\alpha}$. This is also called backward completeness because a dual
form of forward completeness may be considered.  As a very simple
example, let us consider the abstract domain $\mathit{Sign}$
representing the sign of an integer variable, namely $\mathit{Sign} =
\{\bot, \ok{\bZ_{\scriptscriptstyle{\leq 0}}}, 0,
\ok{\bZ_{\scriptscriptstyle{\geq 0}}}, \top\}\in
\Abs(\wp(\bZ)_\subseteq )$. Let us consider the binary concrete operation of
integer addition on sets
of integers, that is 
$X+Y \ok{\ud} \{x+y~|~ x\in X,\, y\in Y\}$, and the square operator on sets
of integers, that is $\ok{X^2} \ok{\ud} \ok{\{x^2 ~|~ x\in X\}}$.  It turns out that the
best correct approximation
$\ok{+^\mathit{Sign}}$ of integer addition in
$\mathit{Sign}$ is sound but not complete~---~because $\alpha(\{-1\} +
\{1\}) = \ok{0 <_{\mathit{Sign}}}  \top = \alpha(\{-1\})
\ok{+^{\mathit{Sign}}} \alpha(\{1\})$~---~ while it is easy to check
that the best correct approximation of
the square operation in $\mathit{Sign}$ is instead complete. 
Let us also recall that backward completeness implies 
fixpoint completeness, meaning that if $\ok{\alpha \circ f = f^\sharp \circ
\alpha}$ then $\alpha(\lfp(f)) = \ok{\lfp(f^\sharp)}$.

A dual form of completeness can be considered.  The
soundness condition $\ok{\alpha \circ f \sqsubseteq f^\sharp\circ
\alpha}$ can be equivalently formulated as $\ok{f\circ \gamma
\sqsubseteq \gamma \circ f^\sharp}$. Forward completeness for $\ok{f^\sharp}$
corresponds to requiring that the equation $\ok{f\circ \gamma = \gamma
\circ f^\sharp}$ holds,
and therefore means that no loss of precision occurs when a concrete computation
$f(\gamma(a))$, for some abstract value $a\in A$, is approximated in
$A$ by $\ok{f^\sharp (a)}$. Let us
notice that backward and forward completeness are orthogonal
concepts. In fact: (1)~we observed above that $\ok{+^\mathit{Sign}}$ is
not backward complete while it is forward complete because for any $a_1,a_2 \in
\mathit{Sign}$, $\gamma(a_1) + \gamma(a_2) = 
\gamma( a_1 \ok{+^\mathit{Sign}} a_2)$: for instance,
$\gamma(\bZ_{\scriptscriptstyle{\geq 0}}) +
\gamma(\bZ_{\scriptscriptstyle{\geq 0}})= \bZ_{\scriptscriptstyle{\geq 0}} = 
\gamma(\bZ_{\scriptscriptstyle{\geq 0}} \ok{+^\mathit{Sign}}
\bZ_{\scriptscriptstyle{\geq 0}})$;  (2)~the 
best correct approximation $\ok{(\cdot)^{2_{\mathit{Sign}}}}$ of
the square operator on $\mathit{Sign}$ is
not forward complete because $\ok{\gamma( \bZ_{\scriptscriptstyle{\geq 0}} )^2} \subsetneq
\gamma(\mathbb{Z}_{\scriptscriptstyle{\geq 0}})=
\gamma( \ok{(\bZ_{\scriptscriptstyle{\geq 0}})^{2_{\mathit{Sign}}}})$
while, as observed above, it is instead backward complete. 

\paragraph{Completeness is an Abstract Domain Property.}
Giacobazzi et al.~\cite{grs00} observed that completeness uniquely
depends upon the abstraction map, i.e.\ upon the abstract domain. This
means that if $\ok{f^\sharp}$ is backward complete for $f$ then the
best correct approximation $\ok{f^A}$ of $f$ in $A$ is backward
complete as well, and, in this case, $\ok{f^\sharp}$ indeed coincides
with $\ok{f^A}$.  Hence, for any abstract domain $A$, one can define a
backward complete abstract operation $\ok{f^\sharp}$ on $A$ if and
only if $\ok{f^A}$ is backward complete. Thus, an abstract domain $A
\in \Abs(C)$ is defined to be backward complete for $f$ iff the
equation $\alpha\circ f = \ok{f^A}\circ \alpha$ holds. This simple
observation makes backward completeness an abstract domain property,
namely an intrinsic characteristic of the abstract domain.  Let us
observe that $\alpha\circ f = \ok{f^A} \circ \alpha$ holds iff $\gamma
\circ \alpha \circ f = \gamma \circ \ok{f^A} \circ \alpha = \gamma \circ \alpha
\circ f \circ \gamma \circ \alpha$ holds, so that $A$ is
backward complete for $f$ when $\mu_A \circ f = \mu_A \circ f \circ
\mu_A$. Thus, a closure $\mu\in \uco(C)$, that defines some abstract
domain, is backward complete for $f$ when $\mu \circ f = \mu \circ f \circ
\mu$ holds.
Analogous observations apply to forward completeness, which is also an
abstract domain property: $A\in \Abs(C)$ is forward complete for $f$
(or forward $f$-complete) when $f \circ \mu_A = \mu_A \circ f \circ
\mu_A$, while a closure $\mu\in \uco(C)$
is forward complete for $f$ when $f \circ \mu = \mu \circ f \circ
\mu$ holds.

\subsection{Shells}\label{shells}
Refinements of abstract domains have been studied from the
beginning of abstract interpretation \cite{CC77,CC79} and led to the
notion of shell of abstract domains \cite{fgr96,GR97,grs00}.   
Given a generic poset $P_\leq$ of semantic objects~---~where $x\leq y$
intuitively means  that $x$ is a ``refinement'' of $y$~---~and a property $\cP
\subseteq P$ of these objects, 
the generic 
notion of \emph{shell} is as follows:
the $\cP$-shell of an object $x \in P$ is defined to be 
an object $s_x \in P$  such that:
\begin{itemize}
\item[{\rm (i)}] $s_x$ satisfies the property $\cP$, 
\item[{\rm (ii)}] $s_x$ is a refinement  of $x$, and
\item[{\rm (iii)}] $s_x$ is the greatest among the objects in $P$ satisfying (i)
and (ii). 
\end{itemize}
Note that if a $\cP$-shell exists then it is unique. Moreover, if the
$\cP$-shell exists for any object in $P$ then it turns out that the
operator that maps any $x\in P$ to its $\cP$-shell is a lower closure
operator on $\cP$, being monotone, idempotent and reductive: this is
called the \emph{$\cP$-shell refinement} operator.  We will be
interested in shells of abstract domains and partitions, namely shells
in the complete lattices of abstract domains and partitions.  Given a
state space $\Sigma$ and a partition property $\cP\subseteq
\Part(\Sigma)$, the $\cP$-shell of $P\in \Part(\Sigma)$ is the
coarsest refinement of $P$ satisfying $\cP$, when this exists.  Also,
given a concrete domain $C$ and a domain property $\cP\subseteq
\Abs(C)$, the $\cP$-shell of $A\in \Abs(C)$, when this exists, is the
most abstract domain that satisfies $\cP$ and refines $A$.  As an
important example, Giacobazzi
et al.~\cite{grs00} constructively showed that
backward complete shells always exist when the concrete functions are
continuous. 

\paragraph{Disjunctive Shells.} 
Consider the abstract domain property
of being disjunctive, namely $\dAbs(C) \subseteq \Abs(C)$. 
As already observed in \cite{CC79}, if $C$ is a completely
distributive lattice\footnote{This roughly means that in $C$ arbitrary glb's distribute
over arbitrary lub's~--~any powerset, ordered w.r.t.\ super-/sub-set
relation, is completely distributive.}  then  
any abstract domain $A\in \Abs(C)$
can be refined to its disjunctive completion $\dc(A)\ud \{\vee_C S~|~ S\subseteq
\gamma(A)\}$. This means that $\dc(A)$ is the most abstract domain
that refines $A$ and is disjunctive, namely it is the disjunctive
shell of $A$. Hence,
the disjunctive shell operator 
$\sd : \Abs(C)\ra \Abs(C)$ is defined as follows:
$$\sd (A ) \ud \sqcup \{ X \in \Abs(C)~|~ X \sqsubseteq A ,\:
X \text{ is disjunctive}\}.$$

\paragraph{Forward Complete Shells.}
Let $F\subseteq\Fun(C)$ (thus functions in $F$ may have any arity) and $S\in
\wp(C)$. We denote by
$F(S) \in \wp (C)$ the image of $F$ on $S$, i.e.\ $F(S) \ud  
\{f(\vec{s})~|~f\in F,\: \vec{s}\in S^{\scriptscriptstyle \ari(f)}\}$, and we say that
$S$ is $F$-closed when $F(S)\subseteq S$. 
An abstract domain 
$A \in \Abs(C)$ is forward $F$-complete when $A$ is forward
complete for any $f\in F$. 
Let us observe that $F$-completeness for an abstract domain $A$ means that
the image $\gamma(A)$ is
closed under the image of functions in $F$, namely  
$F(\gamma(A))\subseteq \gamma(A)$. 
Also note that when $k:C^0\ra C$, i.e.\ $k\in C$ is a constant, $A$ is 
$k$-complete iff $k$ is precisely represented in $A$, i.e.\
$\gamma(\alpha(k)) = k$. 
Let us finally note that any abstract domain is always
forward meet-complete
because any uco is Moore-closed.  

The (forward) $F$-complete shell operator 
$\sS_F : \Abs(C)\ra \Abs(C)$ is defined as follows:
$$\sS_F (A ) \ud \sqcup \{ X \in \Abs(C)~|~ X \sqsubseteq A ,\:
X \text{ is forward $F$-complete}\}.$$
As observed in \cite{gq01,rt06},   
it turns out that for any abstract domain $A$, $\sS_F (A)$ is
forward $F$-complete, namely forward complete shells always exist.
When $C$ is finite, note that for the
meet operator $\wedge:C^2 \ra C$ we have
that, for any $F$, $\sS_F=\sS_{F\cup\{\wedge\}}$, because uco's 
(that is, abstract domains) 
are meet-closed.

A forward complete shell $\sS_F(A)$ is a more concrete abstraction
than $A$. How to characterize $\sS_F(A)$? 
As shown in \cite{rt06}, forward complete shells admit a constructive
fixpoint characterization. Let $\ok{F^{\scriptscriptstyle \cM}}: \Abs(C)\ra \Abs(C)$ be defined as follows:
$\ok{F^{\scriptscriptstyle \cM}}(X)\ud\cM(F(\gamma(X)))$, 
namely $\ok{F^{\scriptscriptstyle \cM}} (X)$ is the most
abstract domain that contains the image of $F$ on $\gamma(X)$. 
Given $A \in \Abs (C)$, we consider the operator
$F_A: \Abs(C)\ra \Abs(C)$ defined by the reduced product 
$F_A (X) \ud  A \sqcap \FM(X)$. Let us observe that
$F_A(X) = \ok{\cM(\gamma(A) \cup
F(\gamma(X)))}$ and that $F_A$ is
monotone and therefore 
admits the greatest
fixpoint which provides the forward $F$-complete shell of $A$:

\begin{equation}\label{clfp}
\sS_F (A) =
\gfp (F_A ).
\end{equation}

\begin{example2}
  \label{exc} \rm
  Let $\Sigma=\{1,2,3,4\}$
  and $R \subseteq \Sigma\times \Sigma$ be the relation $\{(1,2),
(2,3), (3,4), (4,4)\}$.  Let us consider the post transformer 
$\post_R:\wp(\Sigma)\ra\wp(\Sigma)$. Consider the abstract domain $A=\{ \varnothing, 
2,1234\} \in \Abs(\wp(\Sigma)_\subseteq)$.  
We have that 
$\sS_{\post_R}(A)=\{\varnothing,2,3,4, 34,234, 1234\} $ because by \ref{clfp}:
  \begin{align*}
    X_0 &= \{1234\} \mbox{~~~~~(most abstract domain)} \\
    X_1 &= \cM (A \cup \post_R (X_0))=\cM(A \cup \{234\})=
\{\varnothing, 2, 234,1234\} \\
    X_2 &= \cM(A \cup \post_R (X_1)) = \cM(A \cup \{\varnothing,3,34,234\})
    = \{\varnothing, 2, 3, 34, 234, 1234 \} \\
    X_3 &= \cM(A \cup \post_R (X_2)) 
    = \cM( A \cup \{ \varnothing, 3, 4, 34, 234\})
    =\{ \varnothing, 2,3,4, 34, 234, 1234\} \\
   X_4 &= \cM(A \cup \post_R (X_3)) 
        = \cM( A \cup \{ \varnothing, 3, 4, 34, 234\})
=X_3 ~~~~~\mbox{(greatest fixpoint).}~~~~\qed 
\end{align*}
\end{example2}

\section{Generalized Strong Preservation}\label{gsp}
Let us recall from \cite{rt06} how partitions, i.e.\ standard abstract
models, can be viewed as specific abstract domains and how strong
preservation in standard abstract model checking can be cast 
as forward completeness of abstract
interpretations. 

\subsection{Partitions as Abstract Domains}\label{pads}
Let $\Sigma$ be any (possibly infinite) set of system states. 
As shown in \cite{rt06}, 
it turns out that the lattice of state partitions $\Part(\Sigma)$ can be
viewed as an abstraction of the 
lattice of abstract domains $\Abs(\wp(\Sigma))$. 
This is important for our 
goal of performing  an \emph{abstract fixpoint computation} 
on the
abstract lattice of partitions $\Part(\Sigma)$
of a forward
complete shell in $\Abs(\wp(\Sigma))$.

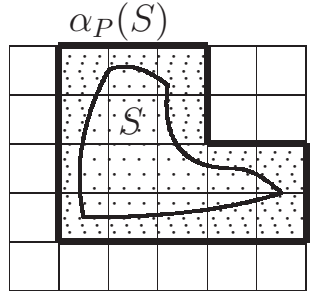
\begin{figure}[t]
\begin{center}
\setlength{\unitlength}{.65cm}
\begin{picture}(6,6)
\linethickness{0.03mm}
\multiput(0,0)(1,0){7}
{\line(0,1){5}}
\multiput(0,0)(0,1){6}
{\line(1,0){6}}
\linethickness{0.35mm}
{
\qbezier(1.5,1.5)(1.2,3)(2,4.5)
\qbezier(2,4.5)(2.5,4.75)(3.2,4.2)
\qbezier(3.2,4.2)(3,2.5)(4.5,2.5)
\qbezier(4.5,2.5)(5,2.5)(5.5,2)
\qbezier(5.5,2)(4,1.5)(1.5,1.5)
\linethickness{0.8mm}
\put(1,1){\line(0,1){4}}
\put(1,5){\line(1,0){3}}
\put(4,5){\line(0,-1){2}}
\put(4,3){\line(1,0){2}}
\put(6,3){\line(0,-1){2}}
\put(1,1){\line(1,0){5}}

\put(2.2,3.2){{\Large $S$}}
\put(1.2,5.3){{\Large $\alpha_P(S)$}}

\multiput(1.15,4.65)(0.3,0){10}{$\cdot$}
\multiput(1.05,4.35)(0.3,0){10}{$\cdot$}
\multiput(1.15,4.05)(0.3,0){10}{$\cdot$}
\multiput(1.05,3.75)(0.3,0){10}{$\cdot$}
\multiput(1.15,3.45)(0.3,0){10}{$\cdot$}
\multiput(1.05,3.15)(0.3,0){10}{$\cdot$}
\multiput(1.15,2.85)(0.3,0){10}{$\cdot$}
\multiput(1.05,2.55)(0.3,0){17}{$\cdot$}
\multiput(1.15,2.25)(0.3,0){16}{$\cdot$}
\multiput(1.05,1.95)(0.3,0){17}{$\cdot$}
\multiput(1.15,1.65)(0.3,0){16}{$\cdot$}
\multiput(1.05,1.35)(0.3,0){17}{$\cdot$}
\multiput(1.15,1.05)(0.3,0){16}{$\cdot$}

\multiput(1.1,4.5)(0.3,0){10}{$\cdot$}
\multiput(1.1,4.2)(0.3,0){3}{$\cdot$}
\multiput(3.2,4.2)(0.3,0){3}{$\cdot$}
\multiput(1.1,3.9)(0.3,0){3}{$\cdot$}
\multiput(3.2,3.9)(0.3,0){3}{$\cdot$}
\multiput(1.1,3.6)(0.3,0){2}{$\cdot$}
\multiput(3.2,3.6)(0.3,0){3}{$\cdot$}
\multiput(1.1,3.3)(0.3,0){2}{$\cdot$}
\multiput(3.2,3.3)(0.3,0){3}{$\cdot$}
\multiput(1.1,3)(0.3,0){2}{$\cdot$}
\multiput(3.5,3)(0.3,0){2}{$\cdot$}
\multiput(1.1,2.7)(0.3,0){1}{$\cdot$}
\multiput(3.5,2.7)(0.3,0){8}{$\cdot$}
\multiput(1.1,2.4)(0.3,0){1}{$\cdot$}
\multiput(4.5,2.4)(0.3,0){5}{$\cdot$}
\multiput(1.1,2.1)(0.3,0){1}{$\cdot$}
\multiput(5.4,2.1)(0.3,0){2}{$\cdot$}
\multiput(1.1,1.8)(0.3,0){1}{$\cdot$}
\multiput(5.4,1.8)(0.3,0){2}{$\cdot$}
\multiput(1.1,1.5)(0.3,0){1}{$\cdot$}
\multiput(4.2,1.5)(0.3,0){6}{$\cdot$}
\multiput(1.1,1.2)(0.3,0){16}{$\cdot$}

}
\end{picture}
\end{center}
\caption{Partitions as abstract domains.}\label{figuco}
\end{figure}

A  partition
$P\in \Part(\Sigma)$ can be viewed as an abstraction of
$\wp(\Sigma)_\subseteq$ as follows:  
any $S\subseteq
\Sigma$ is over approximated by the unique minimal cover of $S$ in $P$, 
namely by the 
union of all the  blocks $B\in P$ such that $B\cap S\neq
\varnothing$. A graphical example is depicted in
Figure~\ref{figuco}. This abstraction is formalized
by a GI $(\alpha_P,\wp(\Sigma)_\subseteq,\wp(P)_\subseteq, \gamma_P)$
where: 
$$\alpha_P(S) \ud \{B\in P~|~ B \cap S\neq \varnothing\}~~~~~~~
\gamma_P(\mathcal{B})\ud \cup_{B\in \mathcal{B}} B.$$
We can therefore define a function $\pad:\Part(\Sigma)\ra
\Abs(\wp(\Sigma))$ that maps 
any partition $P$ to an abstract domain
$\pad(P)$ which is called \emph{partitioning}. In
general, an
abstract domain $A\in \Abs(\wp(\Sigma))$ is called partitioning when $A$ is
equivalent to an abstract domain $\pad(P)$ for some partition $P\in \Part(\Sigma)$. 
Accordingly, a closure $\mu\in \uco(\wp(\Sigma))$ that coincides with 
$\gamma_P \circ \alpha_P$, for some
partition $P$, is called partitioning. 
It can be shown that an abstract domain $A$ 
is partitioning iff its image $\gamma(A)$ is closed under complements, that is,
$\forall S\in \gamma(A).\: \complement (S) \in \gamma(A)$.
We denote by $\Absp(\wp(\Sigma))$ and $\ucop(\wp
(\Sigma ))$  the sets of, respectively, partitioning abstract
domains and closures on $\wp(\Sigma)$.

Partitions can thus be viewed as representations of particular 
abstract domains. On the other hand,
it turns out that abstract domains can be abstracted to partitions. 
An abstract domain $A\in \Abs(\wp(\Sigma)_\subseteq)$ 
induces a state equivalence $\equiv_A$ on $\Sigma$ by
identifying those states that cannot be distinguished by $A$:
$$s \equiv_A s' \text{~~~iff~~~} \alpha(\{s\}) = \alpha(\{s'\}).$$
For any $s\in \Sigma$, $[s]_A \ud \{s'\in \Sigma~|~
\alpha(\{s\}) = \alpha(\{s'\})\}$ is a block of the state partition $\pr(A)$
induced by $A$:
$$\pr(A) \ud \{[s]_A ~|~ s\in \Sigma\}.$$ 
Thus, $\pr:
\Abs(\wp(\Sigma)) \ra \Part(\Sigma)$ is a mapping from abstract domains to
partitions.

\begin{example2}\label{simple}
Let $\Sigma =\{1,2,3,4\}$ and let us specify  abstract domains as
uco's on $\wp(\Sigma)$. The abstract domains
$A_1 = \{\varnothing, 12, 3,4,1234\}$, $A_2 = \{\varnothing, 12,
3, 4, 34, 1234\}$, $A_3 = \{\varnothing, 12, 3, 4, 34, 123, 124,
1234\}$, 
$A_4 = \{ 12,
123, 124, 1234\}$ and $A_5 = \{\varnothing, 12,
123, 124, 1234\}$ all  induce the same
partition $P=\pr(A_i) = \{12,3,4\}\in \Part(\Sigma)$. For example, 
$\alpha_{A_5}(\{1\})=\alpha_{A_5} (\{2\})=\{1,2\}$, $\alpha_{A_5}(\{3\})=\{1,2,3\}$ and
$\alpha_{A_5} (\{4\})=\{1,2,3,4\}$ so that $\pr(A_5)=P$. 
Observe that $A_3$ is the
only partitioning abstract domain because $\pad(P)=A_3$.  \qed
\end{example2}

Abstract domains of $\wp(\Sigma)$ carry
additional information other than the underlying state
partition and this additional information distinguishes
them. As shown in \cite{rt06}, it turns out that this can be precisely stated by
abstract interpretation since the above mappings $\pr$ and $\pad$
allows us to view the whole lattice of partitions of
$\Sigma$ as a (``higher-order'') abstraction of the lattice of abstract
domains of $\wp(\Sigma)$:
\begin{equation*}
(\pr, \Abs(\wp(\Sigma))_\sqsupseteq,\Part(\Sigma)_\succeq,
\pad) \text{ is a GI.} 
\end{equation*}

\noindent
As a consequence, 
the mappings $\pr$ and $\pad$ give rise to an order isomorphism
between state partitions and partitioning abstract domains: 
$\Part(\Sigma)_\preceq \cong \Absp(\wp(\Sigma))_\sqsubseteq$. 
 
\subsection{Abstract Semantics and Generalized Strong Preservation}

\paragraph{Concrete Semantics.}
We consider temporal specification languages $\fL$ whose
state formulae $\varphi$ are inductively defined by: 
$$\fL\ni \varphi ::= p ~|~ f(\varphi_1, ...,\varphi_n) $$
where $p$ ranges over a (typically finite) set of atomic propositions $\AP$, while $f$
ranges over a finite set $\mathit{Op}$ of operators. $\AP$ and $\Op$
are also denoted, respectively, by $\AP_\fL$ and $\Op_\fL$. Each
operator $f\in \mathit{Op}$ has an arity\footnote{It would be
possible to consider generic operators whose arity is any
possibly infinite ordinal, thus allowing, for example, infinite
conjunctions or disjunctions.} 
$\sharp(f) >0$.

Formulae in $\fL$ are interpreted on a \emph{semantic structure} $\cS
= (\Sigma, I)$ where $\Sigma$ is 
any (possibly
infinite) set of states and $I$ is an interpretation function $I: \AP
\cup \mathit{Op} \ra \Fun(\wp(\Sigma))$ that
maps $p\in \AP$ to the set $I(p)\in \wp(\Sigma)$ and $f\in \mathit{Op}$ to 
the function $I(f): \wp(\Sigma)^{\sharp(f)} \ra \wp(\Sigma)$. 
$I(p)$ and $I(f)$ are also denoted by, respectively,
$\boldsymbol{p}$ and $\boldsymbol{f}$. Moreover, $\boldsymbol{AP}\ud
\{\boldsymbol{p} \in \wp(\Sigma)~|~ p \in \AP\}$ and $\boldsymbol{Op}
\ud \{ \boldsymbol{f}:\wp(\Sigma)^{\sharp(f)}\ra \wp(\Sigma)~|~ f\in \Op\}$.
The 
\emph{concrete state semantic function} $\grasse{\cdot}_\cS: \fL\ra
\wp(\Sigma)$ evaluates a formula $\varphi\in \fL$ to the set of states making
$\varphi$ true w.r.t.\ the semantic structure $\cS$:
$$\grasse{p}_\cS = \boldsymbol{p} \mbox{{\rm ~~~and~~~ }}
\grasse{f(\varphi_1,...,\varphi_n)}_\cS =
\boldsymbol{f}(\grasse{\varphi_1}_\cS,...,\grasse{\varphi_n}_\cS).$$
Semantic structures generalize the role of Kripke
structures. In fact, in standard model checking a semantic structure
is usually defined through a Kripke structure $\cK$ so that the
interpretation of logical/temporal operators is defined in terms of
paths in $\cK$ and standard logical operators.
In the following, we freely use standard logical and temporal
operators together with their usual interpretations: for
example, $I(\wedge)= \cap$, $I(\vee)= \cup$, $I(\neg)=
\complement$, and if $\sra$ denotes a transition relation in $\cK$
then $I(\mathrm{EX})= \pre_\sra$, $I(\mathrm{AX})=\pret_\sra$,
etc. 

If $g$ is any operator with arity $\sharp(g)=n>0$, whose interpretation
is given by $\boldsymbol{g}:\wp(\Sigma)^{n} \ra \wp (\Sigma )$, and
$\cS=(\Sigma,I)$ is a semantic structure then we say that a language
$\fL$ is \emph{closed under} $g$ for $\cS$ when for any $\varphi_1,
...,\varphi_{n} \in \fL$ there exists some $\psi\in\fL$ such that
$\boldsymbol{g}(\grasse{\varphi_1}_\cS,...,\grasse{\varphi_{n}}_\cS) =
\grasse{\psi}_\cS$. 
In particular, a language $\fL$ is closed under
(finite) infinite logical conjunction for $\cS$ iff for any (finite) 
$\Phi\subseteq \fL$, there
exists some $\psi\in\fL$ such that $\bigcap_{\varphi\in
\Phi}\grasse{\varphi}_\cS = \grasse{\psi}_\cS$. In particular, let us
note that if $\fL$ is
closed under infinite logical conjunction then it must exist some
$\psi\in \fL$ such that $\cap \varnothing
= \Sigma  = \grasse{\psi}_\cS$, namely $\fL$ is able to express the tautology
$\mathit{true}$. Let us also remark that 
if the state space $\Sigma$ is finite and 
$\fL$ is closed under logical conjunction then we also
mean that there exists some $\psi\in \fL$ such that $\cap \varnothing
= \Sigma  = \grasse{\psi}_\cS$. Finally, note that if $\fL$ is
closed  under negation and (infinite) logical conjunction then $\fL$ is
closed under (infinite) logical disjunction as well.

\paragraph{Abstract Semantics.}

Abstract interpretation allows to define abstract
semantics. Let $\fL$ be a language
and $\cS=(\Sigma,I)$ be a semantic structure for $\fL$.  An
\emph{abstract semantic structure} $\cS^\sharp = (A,I^\sharp)$ is
given by an abstract domain $A\in 
\Abs(\wp(\Sigma)_\subseteq)$ and by an abstract interpretation function
$I^\sharp: \AP \cup \Op \ra \Fun(A)$. An abstract
semantic structure $\cS^\sharp$ therefore induces an \emph{abstract
semantic function} $\grasse{\cdot}_{\cS^\sharp}: \fL\ra A$
that evaluates formulae in $\fL$ to
abstract values in $A$.
In particular, the abstract domain $A$ systematically induces an abstract
semantic structure $\cS^A=(A,I^A)$ where $I^A$ is the best correct
approximation of $I$ on $A$, i.e.\ $I^A$ interprets atoms $p$ and
operators $f$ as best correct approximations on $A$ of,
respectively, 
$\boldsymbol{p}$ and $\boldsymbol{f}$: for any $p\in \AP$ and $f\in \Op$, 
$$I^A (p)\ud \alpha(\boldsymbol{p}) ~~~\text{ and }~~~ I^A (f) \ud
\boldsymbol{f}^A=\alpha\circ \boldsymbol{f}\circ \tuple{\gamma,...,\gamma}.$$
Thus, the abstract domain $A$ always induces an abstract
semantic function $\grasse{\cdot}_{\cS^A}: \fL\ra A$,   
also denoted by $\grasse{\cdot}_\cS^A$, which is therefore defined by:
$$\grasse{p}_\cS^A = \alpha(\boldsymbol{p})
~~~\text{ and }~~~
\grasse{f(\varphi_1,...,\varphi_n)}_\cS^A =
\boldsymbol{f}^A(\grasse{\varphi_1}^A_\cS,...,\grasse{\varphi_n}^A_\cS).$$

\paragraph{Standard Strong Preservation.}

A state semantics $\ok{\grasse{\cdot}_{\cS}}$, for a semantic/Kripke
structure $\cS$, induces a state logical equivalence $\ok{\equiv_\fL^\cS} \,
\subseteq \ok{\Sigma \times \Sigma}$ as usual: 
\begin{equation*}
s\,\ok{\equiv_\fL^\cS}\,
s'  \text{~~~~iff~~~~} \ok{\forall \varphi\in \fL}.\: s\in
\ok{\grasse{\varphi}_{\cS}} \,\Leftrightarrow\, s'
\in \ok{\grasse{\varphi}_{\cS}}.
\end{equation*}
Let
$\ok{P_\fL} \in \ok{\Part(\Sigma)}$ be the partition induced by
$\ok{\equiv^\cS_\fL}$  (the
index $\cS$ denoting the semantic/Kripke structure is omitted).
For a number of well known temporal languages like $\ok{\CTLS}$, $\ok{\ACTLS}$,
$\ok{\CTLSX}$, it turns out that if a partition is more refined than
$P_\fL$ then it induces a standard \emph{strongly preserving} (s.p.)
abstract model. This means that if $\fL$ is interpreted on a Kripke
structure $\cK=(\Sigma, \sra,\ell)$ and $P \preceq P_\fL$ then one can
define an abstract Kripke structure $\cA= (P, \ok{\sra^\sharp},
\ok{\ell^{\sharp}})$ having the partition $P$ as abstract state space 
that strongly preserves $\fL$: 
for any $\varphi \in \fL$, $s\in \Sigma$ and $B\in P$ such
that $s\in B$, we have that $\ok{B \models^\cA \varphi}$ (that is, 
$B \in \ok{\grasse{\varphi}_\cA}$) if and only if $\ok{s \models^\cK
\varphi}$ (that is, $s \in
\ok{\grasse{\varphi}_\cK}$). Let us recall a couple of well-known
examples (see e.g.~\cite{cgp99,dams96}): 
\begin{itemize}
\item[{\rm (i)}] Let
$\ok{P_{\ACTLS}}\in \Part(\Sigma)$ be the partition induced by
$\ACTLS$ on some $\cK=(\Sigma,\sra,\ell)$. 
If $P \preceq \ok{P_{\ACTLS}}$
then the abstract Kripke
structure $\cA = (P, \ok{\sra^{\forall\exists}},\ell_P)$ strongly preserves
$\ACTLS$, where 
$\ell_P (B) =\cup \{\ell(s)~|~s\in B\}$ and
$\ok{\sra^{\forall\exists}}\subseteq P\times P$ is defined as: 
$B_1\,\ok{\sra^{\forall\exists}} \, B_2 \;\,\Lra\;\,\forall
s_1\in B_1.\: \exists s_2\in B_2. \:s_1\sra s_2$.
\item[{\rm (ii)}] Let
$\ok{P_{\CTLS}}\in \Part(\Sigma)$ 
be the partition induced by $\CTLS$ on $\cK$. 
If $P \preceq \ok{P_{\CTLS}}$
then the abstract Kripke
structure $\cA = (P, \ok{\sra^{\exists\exists}},
\ell_P)$ strongly preserves $\CTLS$, where
$B_1\, \ok{\sra^{\exists\exists}}\, B_2  \;\,\Lra\;\,\exists
s_1\in B_1,s_2\in B_2.\:s_1 \sra s_2$.
\end{itemize}

Following Dams \cite[Section~6.1]{dams96} and Henzinger et
al.~\cite[Section~2.2]{hmr05}, the notion of strong preservation can
be given w.r.t.\ a mere state partition rather than w.r.t.\ an
abstract Kripke structure.  A partition $P\in \Part(\Sigma)$ is
strongly preserving\footnote{Dams
\cite{dams96} uses the term ``fine'' instead of ``strongly
preserving''.} for $\fL$
(when interpreted on a semantic/Kripke structure $\cS$) 
if $P \preceq \ok{P_\fL}$. In this sense,
$\ok{P_\fL}$ is the coarsest partition that is strongly preserving for
$\fL$. 
For a number of well known temporal languages, like 
$\ACTLS$, $\CTLS$ (see, respectively, the above points~(i) and (ii)), $\CTLSX$ and
the fragments of the $\mu$-calculus described by Henzinger et al.~\cite{hmr05}, 
it turns out that if $P$ is
strongly preserving for $\fL$ then the abstract
Kripke structure $(P, \ok{\sra^{\exists\exists}},
\ell_P)$ is strongly
preserving for $\fL$. 
In particular,  $(P_\fL, \ok{\sra^{\exists\exists}},
\ell_{P_\fL})$  is strongly preserving for $\fL$ and, additionally, $P_\fL$ is the
smallest possible abstract state space, namely   
if $\cA = (A, \ok{\sra^\sharp}, \ok{\ell^\sharp})$ is an abstract Kripke structure that
strongly preserves $\fL$ then $|P_\fL| \leq |A|$.

\paragraph{Generalized Strong Preservation.}
Intuitively, the partition 
$P_\fL$ is an abstraction of the
state semantics $\grasse{\cdot}_{\cS}$. Let us make this intuition precise.
Following \cite{rt06}, an abstract domain $A \in \Abs(\wp(\Sigma))$
is defined to be strongly
preserving
for $\fL$ (w.r.t.\ $\cS$) when
  for any $S\in \wp(\Sigma)$ and $\varphi \in \fL$:~
$\alpha(S)\leq
  \ok{\grasse{\varphi}_{\cS}^A} \: \Lra \: S\subseteq
  \grasse{\varphi}_{\cS}$. 
This generalizes
strong preservation from partitions to abstract domains 
because, by exploiting
the isomorphism in Section~\ref{pads} 
between partitions and partitioning abstract domains, it turns
out that $P$ is a s.p.\ partition for $\fL$ w.r.t.\ $\cS$ iff
$\pad(P)$ is a s.p.\ abstract domain 
for $\fL$ w.r.t.\ $\cS$. 

\paragraph{Forward Complete Shells and Strong Preservation.} 
Partition refinement
algorithms for computing behavioural equivalences like bisimulation
\cite{pt87}, simulation equivalence \cite{bg03,hhk95,TC01} and (divergence
blind) stuttering equivalence \cite{gv90} are used
in abstract model checking to
compute the coarsest strongly preserving partition of temporal languages
like $\CTLS$ or the
$\mu$-calculus for the case of bisimulation equivalence, $\ACTLS$ for simulation
equivalence and $\CTLSX$ for stuttering equivalence.
Let us recall from \cite{rt06}
how the input/output behaviour of 
these partition refinement algorithms can be generalized through
abstract interpretation. 
Given a language $\fL$ and a concrete state space $\Sigma$, 
partition refinement algorithms 
work by iteratively refining an
initial partition $P$ within the lattice of partitions $\Part(\Sigma)$
until the fixpoint $P_\fL$ is reached.  The input partition $P$
determines a set $\AP_P$ of atoms and a corresponding interpretation $I_P$ as
follows: $\AP_P\ud \{p_B~|~B\in P\}$ and $I_P(p_B) \ud B$.  More in
general, any $\cX\subseteq \wp(\Sigma)$ determines a set
$\{p_X\}_{X\in \cX}$ of atoms with interpretation $I_\cX( p_X) =X$. In
particular, this can be done for an abstract domain $A \in
\Abs(\wp(\Sigma))$ by considering its concretization
$\gamma(A)\subseteq \Sigma$, namely $A$ is viewed as a set of atoms
with interpretation $I_A (a) = \gamma(a)$. Thus, an abstract domain
$A\in
\Abs(\wp(\Sigma))$ 
together with a set of functions $F\subseteq
\Fun(\wp(\Sigma))$ determine a language $\fL_{A,F}$, with atoms in
$A$, operations in $F$ and endowed with a semantic structure
$\cS_{A,F}=(\Sigma,I_A \cup I_F)$ such that  for any $a\in A$, $I_A(a)=\gamma(a)$
and for any $f\in F$, $I_F(f)=f$. 
When $\fL_{A,F}$ is closed under infinite logical
conjunction (for finite state spaces this boils down to closure
under finite conjunction) it turns out that the forward complete shell of $A$ for $F$ 
provides exactly 
the most abstract domain in $\Abs(\wp(\Sigma))$ that refines $A$ and is strongly
preserving for $\fL_{A,F}$ (w.r.t.\ $\cS_{A,F}$):
\begin{equation}\label{main2}
\sS_F(A) = \sqcup \{ X\in \Abs(\wp(\Sigma))~|~ X \sqsubseteq A,\, X
\text{ is s.p.\ for } \fL_{A,F}\}
\end{equation}
In other terms, forward complete shells coincide with strongly
preserving shells. 

On the other hand, let $P_\ell$ denote the state partition induced by the
state labeling of a semantic/Kripke structure and let $\fL$ be 
closed under logical conjunction
and negation. Then, the coarsest s.p.\ partition $P_\fL$ can be
characterized as a 
forward complete shell as follows:
\begin{equation}\label{coro2}
P_\fL = \pr(\sS_{\boldsymbol{Op}_{\fL}} (\pad(P_\ell))).
\end{equation}

\begin{figure}[t]
\begin{center}
  \mbox{\xymatrix{
&&      *++[o][F]{1} \ar@/^/[dll] \ar[dl] \ar[dr] \ar[drr] ^(0.12){p} & &\\
      *++[o][F]{2} \ar@/^/[urr] ^(0.12){p} &  
      *++[o][F]{3} \ar[l] ^(0.27){p} \ar@/_/[rr] & &
      *++[o][F]{4}  \ar@/_/[ll] ^(0.2){q} & 
      *++[o][F]{5} \ar@/^1.25pc/[lll] \ar[l] \ar@(ur,dr)[] ^(0.12){p} \\
    }
  }
\end{center}
\caption{A Kripke structure.}\label{fig1}
\end{figure}
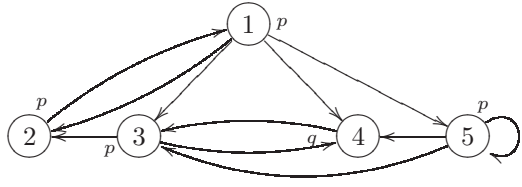

\begin{example2}
Consider the following simple language $\fL$ 
$$
\varphi ::= p ~|~ \varphi_1
\wedge \varphi_2 ~|~
\mathrm{EX}\varphi 
$$
and the Kripke structure $\cK$ depicted in Figure~\ref{fig1}, where
superscripts determine the labeling function $\ell$ and the interpretation of
$\mathrm{EX}$ in $\cK$ is the predecessor operator. 
The labeling function $\ell$ determines the partition
$P_\ell =\{\boldsymbol{p}=1235, \boldsymbol{q}=4\}\in \Part(\Sigma)$, so
that $\pad(P_\ell)= \{\varnothing, 1235,4,12345\}\in \Abs(\wp(\Sigma))$. 
Abstract domains
are Moore-closed so that $\sS_{\boldsymbol{Op}_\fL}=
\sS_{\pre}$. Let us compute
$\sS_{\pre}(\pad(P_\ell))$. 
\begin{align*} 
X_0 & = \pad(P_\ell)= \{\varnothing, 1235,4,12345\}\\[5pt]
X_1 & = X_0 \sqcap \cM(\pre(X_0))= \cM(X_0 \cup \pre (X_0)) \\ 
& =
\cM(\{\varnothing, 1235,4,12345\} \cup \{
\pre(\{4\})=135\}) = \{\varnothing, 135,1235,4,12345\}\\[5pt]
X_2 & = X_1 \sqcap \cM(\pre(X_1))= \cM(X_1 \cup \pre (X_1)) \\ 
& =
\cM(\{\varnothing, 135,1235,4,12345\} \cup \{
\pre(\{135\})= 1245\}) = \{\varnothing, 15,125,135,1235,4,1245,12345\}\\[5pt]
X_3 & = X_2 \text{~~~~(fixpoint)}
\end{align*}
By (\ref{main2}), $X_2$ is the most abstract domain that strongly
preserves $\fL$. Moreover, by (\ref{coro2}), $P_\fL = \pr 
(X_2)= \{15,2,3,4\}$ is the coarsest partition that strongly preserves
$\fL$. Observe that the abstract domain $X_2$ is not partitioning so
that $\pad(P_\fL) \sqsubset \sS_{\pre}(\pad(P_\ell))$. 
\qed
\end{example2}

\section{GPT: A Generalized Paige-Tarjan Refinement Algorithm}\label{npt}
In order to emphasize the ideas leading to our generalized Paige-Tarjan
algorithm, let us first describe how some features of the
Paige-Tarjan algorithm can be
viewed and generalized from an abstract interpretation perspective.

\subsection{A New Perspective of PT}
Consider a finite 
Kripke structure 
$(\Sigma, \sra, \ell)$ over a set $\AP$ of atoms. In the following, 
$\Part(\Sigma)$ and $\pre_{\sra}$ will be more simply denoted by, respectively, 
$\Part$ and $\pre$.  
As a direct consequence of (\ref{main2}), it turns out \cite{rt06} that
the output $\PT(P)$ of
the Paige-Tarjan algorithm on an input partition $P\in \Part $ is the
partitioning abstraction of  
the forward $\{\pre,\complement\}$-complete shell of $\pad(P)$, i.e.\
$$\PT(P)=\pr(\sS_{\{\pre,\complement\}} (\pad(P))).$$ 
Hennessy-Milner logic $\hml$ is inductively
generated by the logical/temporal operators of conjunction,
negation and existential next-time, so that 
$\boldsymbol{Op}_{\scriptscriptstyle \hml}=\{\cap,\complement,\pre\}$. Moreover, as noted
in Section~\ref{shells}, $\sS_{\{\cap,\complement,\pre\}} =
\sS_{\{\complement,\pre\}}$. 
Hence,  by (\ref{coro2}), we observe that $\PT(P)$ computes the
coarsest partition $P_{\scriptscriptstyle \hml}$ that is strongly
preserving for $\hml$.

On the other hand, equation (\ref{clfp}) provides 
a constructive characterization of forward
complete shells, meaning that it provides a na{\"{\i}}ve fixpoint algorithm for
computing a complete shell $\sS_F(A)=\gfp (F_A )$: begin with $X = \{\Sigma\}=
\top_{\Abs(\wp(\Sigma))}$ and iteratively, at each step, compute $F_A (X)$ until
a fixpoint is reached.  This scheme could be in particular applied for
computing $\sS_{\{\pre,\complement\}} (\pad(P))$.  Note however
this na{\"{\i}}ve fixpoint algorithm is far from being
efficient since at each step $F_A (X)$ always re-computes the
images $f(\vec{x})$ that have already been computed at the previous
step (cf.\ Example~\ref{exc}).

In our abstract interpretation
view, $\PT$ is therefore an algorithm that computes 
\begin{center}
\emph{a particular 
abstraction of a particular forward complete shell}.
\end{center} 
Our goal is to analyze the basic steps of
the $\PT$ algorithm 
in order to investigate whether it can be
generalized from an abstract interpretation perspective to  
an
algorithm that computes 
\begin{center}
\emph{a generic abstraction of a generic forward
complete shell}.
\end{center}  
Let us first isolate in our framework
the following  key points concerning the $\PT$ algorithm.

\begin{lemma}\label{punti}
Let $P\in
\Part$ and $S\subseteq \Sigma$.

\begin{itemize}

\item[{\rm (i)}] 
$\ptsplit(S,P) = \pr(\cM (\pad(P) \cup \{\pre (S)\})) =
\pr(\pad(P)\sqcap \cM(\{\pre(S)\}))$. 

\item[{\rm (ii)}] $\ptrefiners (P) = \{ S \in \pad(P)~|~ \pr (\cM
(\pad(P) \cup \{\pre (S)\})) \prec P \}$. 

\item[{\rm (iii)}] $P$ is $\PT$ stable iff $\{ S \in \pad(P)~|~ \pr (\cM
(\pad(P) \cup\{ \pre (S)\})) \prec P \} = \varnothing$. 

\end{itemize}
\end{lemma}
\begin{proof}
(i) By definition, 
$\ptsplit(S,P) = P \curlywedge \{ \pre (S) , \complement
(\pre(S))\}$. Note that $\pr (\cM(\{\pre(S)\})) =
\pr (\{\pre(S),\Sigma\}) = \{ \pre (S) , \complement
(\pre(S))\}$. Finally, observe that 
$\cM (\pad(P) \cup \{\pre (S)\}) = \pad(P) \sqcap  \cM (\{\pre
(S)\})$. Also, since $\pr:\Abs\wp(\Sigma))_\sqsupseteq \ra
\Part(\Sigma)_\succeq$ is a left-adjoint map and therefore it is
additive, it turns out that 
\begin{align*}
\pr(\cM (\pad(P) \cup \{\pre (S)\})) &=\text{~~~~~[by the equation
shown above]}\\
\pr  (\pad(P) \sqcap  \cM (\{\pre
(S)\})) &= \text{~~~~~[by additivity of $\pr$]}\\
\pr(\pad(P)) \curlywedge \pr (\cM (\{\pre
(S)\})) &= \text{~~~~~[since $\pr \circ \pad = \id$]} \\
P \curlywedge \{ \pre (S) , \complement(\pre(S))\}.& 
\end{align*}
Points (ii) and (iii) follow immediately from (i).
\end{proof}

Given any set $S\subseteq \Sigma$, 
consider a
domain refinement operation
$\refine_{\pre}(S,\cdot):\Abs(\wp(\Sigma)) \ra \Abs(\wp(\Sigma))$
defined as
$$\refine_{\pre}(S,A)
  \ud A \sqcap \cM(\{\pre(S)\}) = \cM(\gamma(A) \cup \{\pre(S)\}).$$
Observe that the best correct
approximation of $\refine_{\pre}(S,\cdot)$ on the abstract domain $\Part$
is $\ok{\refine_{\pre}^{\Part}}(S,\cdot)\!:\Part\! \ra\! \Part$
defined as $$\ok{\refine_{\pre}^{\Part}}(S,P) \ud
\pr(\pad(P)\sqcap \cM(\{\pre(S)\})).$$ 
Thus, 
Lemma~\ref{punti}~(i) provides
a characterization of the PT splitting step as best correct
approximation of $\refine_{\pre}$ on $\Part$. 
In turn, Lemma~\ref{punti}~(ii)-(iii) yield 
a characterization of $\ptrefiners$ and $\PT$ stability based 
on this best correct  approximation $\ok{\refine_{\pre}^{\Part}}$.  
As a consequence, $\PT$ may be reformulated as
follows.
\begin{center}
{\small
$
\begin{array}{|l|}
\hline \\[-9pt]
~\mbox{{\bf while~}} \{ T \in \pad(P)~|~
\refine_{\pre}^{\Part} (T,P) 
\prec P \}   \neq \varnothing ~\mbox{{\bf do}}~\\
~~~~~~\mbox{{\bf choose~}} S \in \{ T \in \pad(P)~|~
\refine_{\pre}^{\Part}(T,P) \prec P \}  ;~ \\
~~~~~~P :=\refine^{\Part}_{\pre} (S,P);\\
~\mbox{{\bf endwhile}}\\[1pt]
\hline
\end{array}
$
}
\end{center}
In the following, this view of $\PT$ is generalized to any 
abstract domain in $\Abs(\wp(\Sigma))$ and 
some conditions ensuring the
correctness of this generalized algorithm are isolated.

\subsection{Generalizing PT} 
We generalize Lemma~\ref{punti} as follows.
Let $F\subseteq
\Fun(\wp(\Sigma))$. We define a family of
domain refinement operators 
$\refine_f \!:\! \ok{\wp(\Sigma)^{\scriptscriptstyle \ari{(f)}}} \!\ra\!  
(\Abs(\wp(\Sigma)) \!\ra\! \Abs(\wp(\Sigma)))$ 
indexed on functions $f\in F$ and
tuples of sets $\ok{\vec{S}}\in
\ok{\wp(\Sigma)^{\scriptscriptstyle \ari(f)}}$: 
\begin{itemize}
\item[{\rm (i)}] 
$\refine_f (\ok{\vec{S}},A) \ud A \sqcap \cM (\ok{\{f(\vec{S})\}})$.
\end{itemize}
\noindent
A tuple $\ok{\vec{S}}$ is called a $F$-refiner for an abstract domain 
$A$ when there exists $f\in
F$ such that $\ok{\vec{S}}\in \ok{\gamma(A)^{\scriptscriptstyle \ari(f)}}$ and 
indeed $\ok{\vec{S}}$ may contribute to
refine $A$ w.r.t.\ $f$, i.e., $\ok{\refine_f(\vec{S},A )} \sqsubset
A$. We thus define refiners of an abstract domain as follows:
\begin{itemize}
\item[{\rm (ii)}] $\refiners_f (A ) \ud \{ \ok{\vec{S}}\in \ok{\gamma(A)^{\scriptscriptstyle
\ari(f)}} ~|~ 
\refine_f (\ok{\vec{S}}, A )\sqsubset A \}$;~~~~
$\refiners_F(A)\ud \ok{\cup_{f\in F}} \refiners_f (A)$,
\end{itemize}
and in turn abstract domain stability as follows:
\begin{itemize} 
\item[{\rm (iii)}] $A$ is $F$-stable iff $\refiners_F(A) =\varnothing$.
\end{itemize}

\paragraph{Concrete PT.}
The above observations lead us to design the
following $\PT$-like algorithm
called 
$\CPT_F$ (Concrete $\PT$), parameterized by $F$, which takes as input 
an 
abstract domain $A\in \Abs(\wp(\Sigma))$ and computes the 
forward $F$-complete shell of $A$.
\begin{center}
{\small
$
\begin{array}{|l|}
\hline \\[-9pt]
~\mbox{{\bf input}}\!:~ \text{~abstract~domain~} A \in \Abs(\wp(\Sigma));~ \\
~\mbox{{\bf while~}} (\refiners_F (A ) \neq \varnothing) ~\mbox{{\bf do}}\\
~~~~~~\mbox{{\bf choose~}} \text{~for~some~} f\in F,\; \vec{S} \in \refiners_f(A );~ \\
~~~~~~A :=\refine_f(\vec{S},A);\\
~\mbox{{\bf endwhile}};\\[-2pt]
~\mbox{{\bf output}}\!:~ A;
~~~~~~~~~~~~~~~~~~~~~~~~~~~~~~
~~~~~~~~~~~~~~~~~~\framebox{$\CPT_F$}\mbox{\hspace*{-5pt}}\\[-0.35pt]
\hline
\end{array}
$
}
\end{center}

\begin{lemma}\label{refi}
Let $A\in \Abs(\wp(\Sigma))$.
\begin{itemize}
\item[{\rm (i)}]
$A$ is forward $F$-complete iff $\refiners_F (A )=\varnothing$.  
\item[{\rm (ii)}]
Let $\Sigma$ be finite. Then, $\CPT_F$ always terminates and
$\CPT_F(A)=\sS_F(A)$. 
\end{itemize}
\end{lemma}
\begin{proof}
{\rm (i)} Given $f\in F$, notice that 
$A=\refine_f(\vec{S},A)$ iff $f(\vec{S})\in\gamma(A)$.
Hence, $\Refiners_f(A)=\varnothing$ iff for any
$\vec{S}\in\gamma(A)^{\scriptscriptstyle \ari(f)}$, $f(\vec{S})\in\gamma(A)$, namely, iff
$f(\gamma(A))\subseteq \gamma(A)$ iff $A$ is
forward $f$-complete.
Thus, $\Refiners_F(A)=\varnothing$ iff 
$A$ is forward $F$-complete. 
\\
{\rm (ii)} 
We denote by $X_i\in \uco(\wp(\Sigma))$, $f_i\in F$ and
$\vec{S}_i\in \refiners_{f_i}(\mu_i)$ the sequences of, respectively,
uco's, functions in $F$ and refiners that are iteratively computed in some run
of $\CPT_F(A)$, where $X_0=A$. Observe that $\{X_i\}$ is a decreasing
chain in $\uco(\wp(\Sigma))_\sqsubseteq$, hence, since $\Sigma$ is
assumed to be finite, it turns out that $\{X_i\}$ is finite.  
We denote by $X_{\fin}$ the last uco in the sequence $\{X_i\}$,
i.e., $\CPT_F(A)=X_{\fin}$. Since 
$\Refiners_F(X_{\fin})=\varnothing$, by point~(i), 
$X_{\fin}$ is forward $F$-complete, and therefore, from 
$X_{\fin} \sqsubseteq A$, we obtain  that
$X_{\fin}\sqsubseteq\sS_F(A)$.

\noindent
Let us show, by induction on $i$, that $X_i\sqsupseteq\sS_F(A)$.
\begin{description}

\item[{\rm ~~~$(i=0)$:}]
  Clearly, $X_0=A\sqsupseteq\sS_F(A)$.

\item[{\rm ~~~$(i+1)$:}] 
By inductive
hypothesis and monotonicity of $\refine_{f_i}$, it turns out that 
$X_{i+1} = \refine_{f_i}(\vec{S}_i,X_i)  \sqsupseteq
\refine_{f_i}(\vec{S_i},\sS_F(A))$. Moreover, 
by point~(i), 
since $\sS_F(A)$ is
forward $f$-complete, we have that $\refine_{f_i}(\vec{S_i},\sS_F(A)) = 
\sS_F(A)$. 
\end{description}
Thus, we obtain the thesis $X_{\fin}=\sS_F(A)$. 
\end{proof}

\begin{example2}\label{esdue}
\rm 
Let us illustrate $\CPT$ on the abstract domain $A = \{\varnothing, 2, 1234\}$ of 
Example~\ref{exc}. 
$$
  \begin{array}{llll}
    X_0 &\!\!=\!\!& A = \{\varnothing,2, 1234\}
    &\!\!\! S_0 = \{2\} \in \refiners_{\post_R} (X_0)   \\[5pt]
    X_1 &\!\!=\!\!& \cM (X_0 \cup \{{\post_R} (S_0)\})  & \\
     &\!\!=\!\!&\cM(X_0 \cup \{3\})
    = \{\varnothing, 2, 3,1234\} 
    &\!\!\! S_1 = \{3\} \in \refiners_{\post_R} (X_1)    \\[5pt]
    X_2 &\!\! =\!\!& \cM (X_1 \cup \{{\post_R} (S_1)\}) & \\ 
&\!\! =\!\! &\cM(X_1 \cup \{4\})
    = \{\varnothing, 2, 3,4, 1234\} 
    &\!\!\! S_2 = \{1234\} \in \refiners_{\post_R} (X_2)   \\[5pt]
    X_3 &\!\! =\!\!& \cM (X_2 \cup \{{\post_R} (S_2)\}) & \\
&\!\! =\!\!& \cM(X_2 \cup \{234\})
    = \{\varnothing, 2, 3,4, 234, 1234\} 
    &\!\!\! S_3 = \{234\} \in \refiners_{\post_R} (X_3)  \\[5pt]
    X_4 &\!\! =\!\!& \cM (X_3 \cup \{{\post_R} (S_3)\})& \\
&\!\! =\!\!& \cM(X_3 \cup \{34\})
    = \{\varnothing, 2, 3,4, 34, 234, 1234\} 
    &~~\Ra~~\refiners_{\post_R}(X_4) = \varnothing
  \end{array}
$$
Let us note that while in Example~\ref{exc} each step consists in computing the
images of $\post_R$ for the sets belonging to the whole domain at the previous
step and this gives rise to
re-computations, here instead an image $f(S_i)$ is never computed twice because
at each step we nondeterministically choose a refiner $S$ and apply
$\post_R$ to $S$.  
\qed 
\end{example2}

\paragraph{Abstract PT.}
Our goal is to design an abstract version of $\CPT_F$ that works on
a generic abstraction $\cA$ of the lattice of abstract domains $\Abs(\wp(\Sigma))$. 
As recalled in Section~\ref{pads}, partitions can be viewed as a ``higher-order''
abstraction of abstract domains through the Galois insertion 
$(\pr, \Abs(\wp(\Sigma))_\sqsupseteq,\Part(\Sigma)_\succeq,
\pad)$. This is a dual GI since both ordering relations in $\Abs(\wp(\Sigma))$ and
$\Part(\Sigma)$ are reversed. This depends on the fact that we
want to obtain a complete approximation of a forward complete shell, which, by
(\ref{clfp}), is a greatest fixpoint so that we need to approximate
a greatest fixpoint computation ``from above'' instead of ``from below'' as it
happens for a least fixpoint computation.
We thus consider a
Galois insertion $(\alpha, \Abs(\wp(\Sigma))_\sqsupseteq, \cA_\geq,
\gamma)$ of an abstract domain $\cA_\geq$ into the dual lattice of
abstract domains $\Abs(\wp(\Sigma))_\sqsupseteq$. 
The ordering relation of the abstract domain $\cA$ is denoted by $\geq$ because this
makes concrete and abstract ordering notations uniform. 
It is worth remarking that since we require a Galois insertion of $\cA$ into
the complete lattice $\Abs(\wp(\Sigma))$, by standard results \cite{CC79}, 
$\cA$ must necessarily be a complete lattice as well. 
For any $f\in F$, the best
correct approximation
$\ok{\refine_f^\cA}:\ok{\wp(\Sigma)^{\scriptscriptstyle \ari(f)}}\!\ra\! (\cA\!\ra\!
\cA)$ of $\refine_f$  on $\cA$ is therefore defined as usual by:
\begin{itemize}
\item[{\rm (i)}] 
$\ok{\refine^\cA_f(\vec{S}, a)} \ud \alpha(\refine_f (\ok{\vec{S}},\gamma(a)))$.
\end{itemize}
\noindent
Accordingly, 
abstract refiners and stability are defined as follows:
\begin{itemize}
\item[{\rm (ii)}] $\ok{\refiners^\cA_f(a)} \ud \{ \ok{\vec{S}} \in
\ok{\gamma(a)^{\scriptscriptstyle \ari(f)}}~|~
\ok{\refine_f^\cA(\vec{S},a)} < a\}$;~~~~
$\ok{\refiners^\cA_F(a)}\ud \ok{\cup_{f\in F}} \ok{\refiners^\cA_f(a)}$.
\item[{\rm (iii)}] An abstract object $a\in \cA$ is $F$-stable iff
$\ok{\refiners_F^\cA(a)}=\varnothing$.  
\end{itemize}

We may  now define the following abstract version of the above algorithm
$\ok{\CPT_F}$, called $\ok{\GPT^\cA_F}$ (Generalized $\PT$), that is
parameterized on the abstract domain $\cA$. 
\begin{center}
{\small
$
\begin{array}{|l|}
\hline \\[-9pt]
~\mbox{{\bf input}}\!:~ \text{~abstract~object~} a \in \cA;~ \\
~\mbox{{\bf while~}} (\refiners^\cA_F (a ) \neq \varnothing) ~\mbox{{\bf do}}~\\
~~~~~~\mbox{{\bf choose~}} \text{~for~some~} f\in F,\; 
\vec{S} \in \refiners^\cA_f(a );~ \\
~~~~~~a :=\refine^\cA_f(\vec{S},a);\\
~\mbox{{\bf endwhile}};\\[-4pt]
~\mbox{{\bf output}}\!:~ a; 
~~~~~~~~~~~~~~~~~~~~~~~~~~~~
~~~~~~~~~~~~~~~~~~~~\framebox{$\GPT^\cA_F$}\mbox{\hspace*{-5pt}}\\[-0.35pt]
\hline
\end{array}
$
}
\end{center}

$\ok{\GPT^\cA_F(a)}$ computes a
sequence of abstract objects $\ok{\{a_i\}_{i\in \mathbb{N}}}$  
which is a decreasing chain in $\cA_\leq$, namely $a_{i+1} < a_i$.
Thus, in order to ensure termination of $\ok{\GPT_F^\cA}$ it is
enough to consider an abstract domain $\cA$ such that
$\tuple{\cA,\leq}$ satisfies the
descending chain condition (DCC), i.e., every descending chain
is eventually stationary. 
Furthermore, let us remark that 
correctness for $\ok{\GPT^\cA_F}$ means that for any input object $a\in \cA$,
$\GPT^\cA_F(a)$ computes exactly the abstraction in $\cA$ of the
forward $F$-complete shell of the abstract domain $\gamma(a)$, that is
$$\GPT^\cA_F(a)=\alpha (\sS_F(\gamma(a))).$$ 
Note that, by \ref{clfp}, 
$\alpha(\sS_F(\gamma(a))) = \alpha(\gfp(F_{\gamma(a)}))$. 
It should be clear that correctness for $\GPT$ is somehow related to
backward completeness in abstract interpretation. In fact, if the
abstract domain $\cA$ is backward complete for $F_{\gamma(a)} = \lambda X.\gamma(a)
\sqcap \ok{\FM(X)}$ then it is also fixpoint complete (cf.\
Section~\ref{secco}), so that
$\ok{\alpha (\gfp(F_{\gamma(a)}))}= \ok{\gfp(F_{\gamma(a)}^\cA)}$, where
$\ok{F_{\gamma(a)}^\cA}$ is the best correct approximation of the operator 
$\ok{F_{\gamma(a)}}$ on the abstract domain $\cA$.
The intuition is that $\ok{\GPT_F^\cA(a)}$ is an algorithm for
computing the greatest fixpoint $\ok{\gfp(F_{\gamma(a)}^\cA)}$.
Indeed, the following result shows that $\ok{\GPT^\cA_F}$ is
correct when $\cA$ is backward complete for $\FM$, because this implies that
$\cA$ is backward complete for $F_A$, for any abstract domain $A$. Moreover, we
also isolate the following condition ensuring correctness for
$\ok{\GPT^\cA_F}$: the forward $F$-complete shell operator $\sS_F$ maps
domains in $\cA$ into domains in $\cA$, namely
the higher-order abstraction $\cA$ is forward complete for the forward $F$-complete
shell $\sS_F$.

\begin{theorem}\label{main} 
Let $\cA_\leq$ be DCC and assume that one of the following conditions
holds:
\begin{itemize}
\item[{\rm (i)}] $\cA$ is backward complete for $\FM$.
\item[{\rm (ii)}]  $\cA$ is forward complete for $\sS_F$.
\end{itemize}
Then, $\ok{\GPT^\cA_F}$ always terminates and 
for any $a\in \cA$, $\ok{\GPT^\cA_F(a)}=\alpha(\sS_F(\gamma(a)))$. 
\end{theorem}
\begin{proof}
Let us first show the following two facts. For any $a\in \cA$:
\begin{itemize}
\item[{\rm (A)}]  $\Refiners_F(\gamma(a))=\Refiners_F^\cA(a)$.
\item[{\rm (B)}] $\gamma(a)$ is forward $F$-complete iff
$\refiners^\cA_F(a)=\varnothing$. 
\end{itemize}
\noindent
{\rm (A)} Let $f\in F$.   Note that 
  $\refine_f(\vec{S},\gamma(a))= 
\gamma(a)\sqcap\cM(\{\ok{f(\vec{S})}\})$ and therefore
  $\ok{\refine_f^\cA(\vec{S},a)}=\alpha( \gamma(a)\sqcap\ok{\cM(\{f(\vec{S})\}))}
= \alpha(\gamma(a))\wedge_A \ok{\alpha(\cM(\{f(\vec{S})\}))}
= a\wedge_A \ok{\alpha(\cM(\{f(\vec{S})\}))}$.
  Consequently, $\ok{\vec{S}}\in\Refiners_f(\gamma(a))$ iff 
  $\ok{\vec{S}}\in\ok{\gamma(a)^{\scriptscriptstyle \ari(f)}}$ and
  $\ok{\cM(\{f(\vec{S})\})}\not\sqsupseteq \gamma(a)$.
 Likewise,
$\ok{\vec{S}}\in\ok{\Refiners_f^\cA(a)}$ iff 
  $\ok{\vec{S}}\in\ok{\gamma(a)^{\scriptscriptstyle \ari(f)}}$ and
  $\alpha(\cM(\{f(\ok{\vec{S}})\}))\not\geq a$. These are equivalent
properties, because, by Galois insertion, we have that
$\alpha(\cM(\{f(\ok{\vec{S}})\}))\geq a$ iff
  $\cM(\{f(\ok{\vec{S}})\})\sqsupseteq\gamma(a)$.
\\
\noindent
{\rm (B)} $\gamma(a)$ is forward $F$-complete iff 
$\Refiners_F(\gamma(a))=\varnothing$ iff
$\Refiners_F^\cA(a)=\varnothing$, by point~(A).
\\
\noindent
Let us now prove the main result. We denote by  $a_i\in \cA$, $f_i\in F$ and $\vec{S_i}\in
\ok{\Refiners_{f_i}^\cA(a_i)}$ the sequences of, respectively, 
abstract ojects, functions in $F$ and refiners iteratively computed by 
some run of $\ok{\GPT_F^\cA(a)}$, where $a_0=a$. Since $\{a_i\}$ is a decreasing chain
in the abstract domain $\cA_\leq$ which is assumed to be
DCC, it turns out that these sequences are finite. We denote by
$a_{\mathit{fin}}$ the last element in the sequence of $a_i$'s, i.e.,
$\ok{\GPT_F^\cA(a)}=a_{\mathit{fin}}$. Moreover,
we also consider the following sequence of abstract domains:
$X_i \ud \gamma(a_i)\sqcap \FM(\gamma(a_i)) =
\cM(\gamma(a_i) \cup F(\gamma(a_i)))$. Let us notice
that, since $a_{i+1} \leq a_i$, by monotonicity, we have that
$X_{i+1} \sqsubseteq X_i$. Moreover, since
$\ok{\refiners_F^\cA(a_{\mathit{fin}})}=\varnothing$, by point~(B),
$\gamma(a_{\mathit{fin}})$ is forward $F$-complete, hence
$\gamma(a_{\mathit{fin}}) \sqsubseteq \FM(\gamma(a_{\mathit{fin}}))$,
so that
$X_{\mathit{fin}}=\gamma(a_{\mathit{fin}})$. We show that 
$\alpha(X_{\mathit{fin}})=\alpha(\sS_F(\gamma(a)))$, so that 
$a_{\fin} = \alpha(\gamma(a_{\mathit{fin}})) = \alpha(X_{\mathit{fin}})= 
\alpha(\sS_F(\gamma(a)))$ follows.   
By point~(A), 
$\Refiners_F(\gamma(a_{\mathit{fin}}))=\ok{\Refiners_F^\cA}(a_{\mathit{fin}})=\varnothing$,
thus, by Lemma~\ref{refi}~(i),
$\gamma(a_{\mathit{fin}})$  is forward $F$-complete.  
Moreover, $\gamma(a_{\mathit{fin}}) \sqsubseteq \gamma(a_0)=\gamma(a)$ and consequently 
$\gamma(a_{\mathit{fin}}) \sqsubseteq\sS_F(\gamma(a))$.
Hence,  $\alpha(X_{\mathit{fin}}) =\alpha (\gamma(
a_{\mathit{fin}})) \leq \alpha(\sS_F(\gamma(a)))$. 
Let us now show, by induction on $i$, that $\alpha(X_i) 
\geq \alpha(\sS_F(\gamma(a)))$.\\[5pt]
{\rm ~~~$(i=0)$:}
  $X_0 =\gamma(a_0)\sqcap \FM (\gamma(a_0)) = \gamma(a)\sqcap
\FM(\gamma(a))$, hence, since $\sS_F(\gamma(a))\sqsubseteq \gamma(a),
\FM(\gamma(a))$, we have that $\sS_F(\gamma(a))\sqsubseteq X_0$, and
therefore $\alpha(\sS_F(\gamma(a)))\leq \alpha(X_0)$. \\[5pt]
{\rm ~~~$(i+1)$:} Since 
$a_{i+1} =
\alpha(\cM(\gamma(a_i)\cup \{f_i (\vec{S_i}) \} ))$, where
$\vec{S_i}\in \gamma(a_i)$, we have that $f_i(\vec{S_i}) \in \FM
(\gamma(a_i))$. Hence, $\cM(\gamma(a_i)\cup \{f_i (\vec{S_i})\}) \subseteq
\cM(\gamma(a_i)\cup \FM(\gamma(a_i))) = \gamma(a_i)\sqcap
\FM(\gamma(a_i)) = X_i$, namely $X_i \sqsubseteq
\cM(\gamma(a_i)\cup \{f_i (\vec{S_i})\})$, so that $\alpha(X_i) \leq
a_{i+i}$ and $\gamma(\alpha(X_i)) \sqsubseteq \gamma(a_{i+1})$. Moreover:
  \begin{align*}
    \alpha(X_{i+1}) &=  \\ 
    \alpha(\gamma(a_{i+1})\sqcap \FM(\gamma(a_{i+1}))) &=
    \text{~~~~~~[since $\alpha$ is co-additive]}\\
    \alpha(\gamma(a_{i+1})) \sqcap \alpha(\FM(\gamma(a_{i+1})))  &\geq
    \text{~~~~~~[since $\gamma(a_{i+1})\sqsupseteq \gamma(\alpha(X_i))$]} \\
    \alpha(\gamma(\alpha (X_i))) \sqcap \alpha(\FM (\gamma(\alpha(X_i)))) & \geq
    \text{~~~~~~[by induction]}\\ 
    \alpha(\gamma(\alpha (\sS_F (\gamma(a))))) \sqcap
    \alpha(\FM(\gamma(\alpha(\sS_F(\gamma(a)))))) &=
    \text{~~~~~~[since $\alpha \circ \gamma \circ \alpha = \alpha$]}\\
    \alpha(\sS_F (\gamma(a))) \sqcap
    \alpha(\gamma(\alpha(\FM(\gamma(\alpha(\sS_F(\gamma(a)))))))).&  
  \end{align*}

\noindent
Now, both conditions~(i) and~(ii) imply that 
$$\alpha(\gamma(\alpha(\FM(\gamma(\alpha(\sS_F(\gamma(a)))))))) =
\alpha(\gamma(\alpha(\FM(\sS_F(\gamma(a)))))).$$
Thus, we may proceed as follows:
\begin{align*}
\alpha(\sS_F (\gamma(a))) \sqcap
\alpha(\gamma(\alpha(\FM(\rho_A(\sS_F(\gamma(a))))))) &= 
\text{~~~~~~[by either condition~(i) or (ii)]} \\
\alpha(\sS_F (\gamma(a))) \sqcap
\alpha(\gamma(\alpha(\FM(\sS_F(\gamma(a)))))) &= 
\text{~~~~~~[since $\alpha \circ \gamma\circ \alpha = \alpha$]} \\
\alpha(\sS_F (\gamma(a))) \sqcap \alpha(\FM(\sS_F(\gamma(a)))) &= 
\text{~~~~~~[as $\sS_F(\gamma(a))$ is forward $F$-complete]} \\
\alpha(\sS_F (\gamma(a))) \sqcap \alpha(\sS_F(\gamma(a))) &= \\  
\alpha(\sS_F(\gamma(a))). &
\end{align*}

\medskip
\noindent
Thus, this closes the proof. 
\end{proof}

\begin{corollary}\label{cmain} 
Under the hypotheses of Theorem~\ref{main}, 
for any $a\in \cA$, $\ok{\GPT^\cA_F(a)}$ is the $F$-stable shell of $a$. 
\end{corollary}
\begin{proof}
By Theorem~\ref{main}, $\ok{\GPT_F^\cA(a)} \leq a$ and is $F$-stable. Let us
show that $\ok{\GPT_F^\cA(a)}$ indeed is the $F$-stable shell of $a$.  
Let $b\in A$ such that $b\leq a$ and $\ok{\Refiners_F^\cA(b)}=\varnothing$.
Since $b\leq a$, we have that $\gamma(b)\sqsubseteq\gamma(a)$. Moreover, by 
point~(A) in the proof of Theorem~\ref{main}, 
$\ok{\Refiners_F(\gamma(b))}=\ok{\Refiners_F^\cA(b)}=\varnothing$,
so that $\gamma(b)$ is forward $F$-complete by Lemma~\ref{refi}~(i).
Hence, $\gamma(b)\sqsubseteq\sS_F(\gamma(a))$ and thus, by Theorem~\ref{main},
$b=\alpha(\gamma(b))\leq \alpha(\sS_F(\gamma(a))) = \ok{\GPT_F^\cA(a)}$.
\end{proof}

\begin{example2}\label{estre}
\rm
Let us consider again Example~\ref{exc} and~\ref{esdue}. Recall from
Section~\ref{shells}
that the disjunctive shell $\sd:\Abs(\wp(\Sigma)) \ra
\dAbs(\wp(\Sigma))$ maps any abstract domain $A$ to its disjunctive
completion  $\sd (A)=\{\cup S~|~S\subseteq \gamma(A)\}$.
It turns out that the disjunctive shell $\sd$ allows to view
$\dAbs(\wp(\Sigma))_\sqsupseteq$ as an abstraction of
$\Abs(\wp(\Sigma))_\sqsupseteq$, namely $(\sd,
\Abs(\wp(\Sigma))_\sqsupseteq, \dAbs(\wp(\Sigma))_\sqsupseteq, id)$ is
a GI. This is a consequence of the fact that disjunctive abstract
domains are closed under lub's in $\Abs(\wp(\Sigma))$ and therefore 
 $\dAbs(\wp(\Sigma))_\sqsupseteq$ is a Moore-family of
$\Abs(\wp(\Sigma))_\sqsupseteq$. 

\noindent
It turns out that condition~(i) of Theorem~\ref{main} is satisfied
for this GI. In fact, by exploiting the fact that 
$\post_R:\wp(\Sigma)\ra\wp(\Sigma)$ is additive, it is not hard to
verify that $\sd \circ \post_R^{\scriptscriptstyle\cM} \circ \sd = \sd
\circ \post_R^{\scriptscriptstyle\cM}$.  
Thus, let us apply $\ok{\GPT_{\post_R}^{\dAbs}}$ 
to the disjunctive abstract domain $X_0 =
\{\varnothing, 2, 1234\}=
\sd (\{2,1234\})\in \dAbs(\wp(\Sigma))$. 
  \begin{align*}
    &~~~~~~X_0\!\!\!\!\!\!\!\!\!\!\!\!\!\!\! &=\;\;& \{\varnothing, 2, 1234\} 
    &\! \! S_0 = \{2\} \!\in \!\refiners_{\post_R}^{\dAbs} (X_0)   \\[5pt]
    &~~~~~~X_1\!\!\!\!\!\!\!\!\!\!\!\!\!\!\! &=\;\;& \sd (\cM (X_0 \cup \{\post_R (S_0)\})) & \\
    &&\!\!\!\!\!\!=\;\;&\sd (\{\varnothing, 2, 3,1234\}) &\\
    &&\!\!\!\!\!\!=\;\;& \{\varnothing, 2, 3, 23, 1234\} 
    &\! \! S_1 = \{3\} \!\in \!\refiners_{\post_R}^{\dAbs} (X_1)   \\[5pt]
    &~~~~~~X_2\!\!\!\!\!\!\!\!\!\!\!\!\!\!\! & =\;\;& \sd ( \cM (X_1 \cup \{\post_R (S_1)\})) & \\ 
&&=\;\;& \sd
(\{\varnothing, 2, 3,23, 4, 1234\}) & \\
&&=\;\;& \{\varnothing, 2,3,4,23,24, 34, 234, 1234\}  
    &\!\!\!\Ra~~\refiners_{\post_R}^{\dAbs} (X_2)=\varnothing
  \end{align*}
{}From Example~\ref{esdue} we know that $\sS_{\post_R} (X_0)=\{\varnothing,
2,3,4, 34, 234, 1234\}$.
Thus, as expected from Theorem~\ref{main}, 
$\ok{\GPT_{\post_R}^{\dAbs}} (X_0)$ coincides with $\sd (\sS_{\post_R} (X_0))=
\{\varnothing, 2,3,4,$ $23,24,34,$ $234,1234\}$. Note that 
the
abstract fixpoint  has been reached in two iterations, whereas in Example~\ref{esdue} 
the concrete computation by $\CPT_{\post_R}$ needed four iterations. \qed 
\end{example2}

\subsection{An Optimization of GPT}
As pointed out by Paige and Tarjan \cite{pt87}, the $\PT$ algorithm 
works even if splitters are chosen among blocks instead of unions of
blocks, i.e., if $\PTrefiners(P)$ is replaced with 
the subset of ``block refiners''
$\ptblockrefiners(P)\ud\ptrefiners(P)\cap P$. This can be easily
generalized as follows. Given $g\in F$, for any $a\in \cA$, let
$\ok{\subrefiners_g^\cA(a)}\subseteq \ok{\refiners_g^{\cA}(a)}$ be any
subset of refiners. We denote by
$\ok{\IGPT_F^\cA}$ (which stands for Improved $\GPT$) the version of $\ok{\GPT_F^\cA}$ where
$\ok{\refiners_g^\cA}$ is replaced with $\ok{\subrefiners_g^\cA}$. 
If stability for subrefiners is equivalent to stability for refiners
then $\IGPT$ results to be  correct.
\begin{corollary}\label{igpt}
Let $g\in F$ be such that, for any $a\in \cA$,
$\ok{\subrefiners_g^{\cA}(a)} = 
\varnothing$ $\Lra$   $\ok{\refiners_g^{\cA}(a)}= \varnothing$. Then, for any
$a\in \cA$, 
$\ok{\GPT_F^{\cA}(a)} = \ok{\IGPT_F^\cA(a)}$. 
\end{corollary}
\begin{proof}
Let $\ok{\subrefiners_F^\cA(a)}= 
\ok{\subrefiners_g^\cA(a)} \cup
\ok{(\cup_{F\ni f \neq g} \refiners_f^\cA(a))}$.
By hypothesis, we have that $\ok{\subrefiners_F^\cA(a)}\neq \varnothing$ iff
$\ok{\refiners_F^\cA(a)}\neq \varnothing$. Let $\{a_i\}$ be the finite
decreasing chain
of abstract objects computed by $\ok{\IGPT_F^\cA(a)}$. 
Since
$\ok{\subrefiners_F^\cA (\IGPT_F^\cA(a))}=\varnothing$ we have that
$\ok{\refiners_F^\cA(\IGPT_F^\cA(a))} = \varnothing$.
Moreover, since, for any $i$,
$\ok{\subrefiners_g^\cA(a_i)} \subseteq \ok{\refiners_g^\cA(a_i)}$, there exists a
run of $\ok{\GPT_F^\cA(a)}$ which exactly computes the sequence $\{a_i\}$, so
that, by Theorem~\ref{main},
$\ok{\IGPT_F^\cA(a)}=\ok{\GPT_F^\cA(a)}$.  
\end{proof}

\subsection{Instantiating GPT with Partitions}\label{igptp}
Let us now show how the above $\GPT$ algorithm can be instantiated to
the lattice of partitions.  
Assume that the state space $\Sigma$ is finite. 
Recall from Section~\ref{gsp} that the lattice of partitions can be viewed as an
approximation of the lattice of abstract domains through the GI 
$(\pr, \Abs(\wp(\Sigma))_\sqsupseteq,\Part(\Sigma)_\succeq,
\pad)$. 
The following 
properties~(1) and~(2) are  consequences of the fact that a partitioning abstract
domain 
$\pad(P)$ is closed
under complements, i.e.\ $X\in \pad(P)$ iff $\complement(X)\in \pad(P)$. 
\begin{itemize}
\item[{\rm (1)}] 
$\refiners_{\scriptscriptstyle \complement}^{\scriptscriptstyle \Part}(P) = \varnothing$.
\item[{\rm (2)}]  For any $f$ and $\vec{S}\in
\wp(\Sigma)^{\scriptscriptstyle \ari(f)}$,
$\refine_f^{\scriptscriptstyle \Part}(\vec{S},P)=P\curlywedge \{f(\vec{S}),\complement
(f(\vec{S}))\}$. 
\end{itemize}
\noindent
Thus, by Point~(1),  for any $F\subseteq \Fun(\wp(\Sigma))$, a partition
$P\in \Part(\Sigma)$ is $F$-stable iff $P$ is 
$(F \cup\{\complement\})$-stable, 
that is complements can be left out. 
Hence, if $F^{\mbox{{\rm -}}{\scriptscriptstyle\complement}}$ denotes $F\smallsetminus
\{\complement\}$ then $\ok{\GPT^{\Part}_F}$ may be
simplified as follows. 
\begin{center}
{\small
$
\begin{array}{|l|}
\hline \\[-9pt]
~\mbox{{\bf input}}\!:~ \text{~partition~} P \in \Part(\Sigma);~~ \\
~\mbox{{\bf while~}} (\refiners^{\Part}_{F^{\mbox{{\rm -}}{\scriptscriptstyle\complement}}} (P) \neq \varnothing) ~\mbox{{\bf do}}~~\\
~~~~~~\mbox{{\bf choose~}} \mathrm{~for~some~} f\in F^{\mbox{{\rm -}}{\scriptscriptstyle\complement}},\; 
\vec{S} \in \refiners^{\Part}_f(P );~ \\
~~~~~~P := P \curlywedge \{ f(\vec{S}),\complement (f(\vec{S}))\};\\
~\mbox{{\bf endwhile}}\\[-3.5pt]
~\mbox{{\bf output}}\!:~ P;
~~~~~~~~~~~~~~~~~~~~~~~~~~~~~~~~
~~~~~~~~~~~~~~~~~~~~\framebox{$\GPT^{\Part}_F$}\mbox{\hspace*{-5.1pt}}\\[-0.35pt]
\hline
\end{array}
$
}
\end{center}
Note that the number of iterations of
$\ok{\GPT_F^{\Part}}$  is bounded 
by the
height of the lattice $\Part(\Sigma)$, namely by the number of states
$|\Sigma |$. Thus, if
each refinement step involving some $f\in F$ 
takes $O(\mathrm{cost}(f))$ time then the
time complexity of $\GPT_F^{\Part}$ is bounded by $O(|\Sigma|
\max(\{\mathrm{cost}(f)~|~f\in F\}))$.  

Let us now consider a language $\fL$ 
and a semantic structure
$(\Sigma,I)$ for $\fL$.
If $\fL$ is closed under logical conjunction and negation then, for any $A\in
\Abs(\wp(\Sigma))$, $\sS_{\boldsymbol{Op}_{\fL}} (A)$ is closed
under complements and therefore it is a partitioning abstract domain. Thus,
condition~(ii) of Theorem~\ref{main} is satisfied since
$\sS_{\boldsymbol{Op}_{\fL}}$ maps partitioning abstract domains into
partitioning abstract domains.  The following characterization is thus
obtained as a
consequence of (\ref{coro2}).
\begin{corollary}\label{partmain}
$\!\!$If $\fL$ is closed under conjunction and negation then
$\GPT^{\Part}_{\boldsymbol{Op}_\fL} (P_\ell) \!=\! P_\fL$.
\end{corollary}
This provides an algorithm parameterized on a language $\fL$ that
includes propositional logic
for computing the coarsest strongly
preserving partition $P_\fL$.

\paragraph{PT as an Instance of GPT.}
It is now immediate to obtain $\PT$ as an instance of $\GPT$. 
We know that
$\ok{\GPT^{\Part}_{\{\pre ,{\scriptscriptstyle \complement}\}}} 
= \ok{\GPT^{\Part}_{\pre}}$. Moreover, by
Lemma~\ref{punti}~(i)-(ii):
$$ P \curlywedge
\{\pre(S),\complement(\pre(S))\}  = 
\ptsplit(S, P) \mbox{{\rm ~~~and~~~}}
\Refiners_{
\pre}^{ \Part} (P) = \PTrefiners(P).$$
Hence, by Lemma~\ref{punti}~(iii), it turns out that $P\in \Part(\Sigma)$ is $\PT$ stable
iff $\ok{\Refiners_{ \pre}^{\Part}(P)}=\varnothing$. Thus, the instance
$\ok{\GPT_{\pre}^{\Part}}$ provides
\emph{exactly} the $\PT$ algorithm. Also, 
correctness follows from Corollaries~\ref{cmain} and~\ref{partmain}: 
$\ok{\GPT_{\pre}^{\Part} (P)}$ is both the coarsest $\PT$
stable refinement of $P$ and the coarsest strongly
preserving partition $P_{\scriptscriptstyle \hml}$.

\section{Applications}

\subsection{Stuttering Equivalence and Groote-Vaandrager Algorithm}
Lamport's criticism \cite{lam83} of the next-time operator $\mathrm{X}$
in $\CTL$/$\CTLS$ is well known. This motivated the study of  temporal
logics like $\CTLX$/$\CTLSX$ 
obtained  from $\CTL$/$\CTLS$ by removing the next-time operator and led to
study a notion of behavioural \emph{stuttering}-based equivalence
\cite{bcg88,dnv95,gv90}. We are interested here in
\emph{divergence blind stuttering} (dbs for short) equivalence. 
Let $\cK = (\Sigma ,\sra ,\ell)$ be a Kripke structure over a set
$\AP$ of atoms. 
A
relation $R \subseteq \Sigma \times \Sigma $ is a divergence blind 
stuttering relation on $\cK$ if for any
$s,s'\in \Sigma $ such that $sR s'$: 
\begin{itemize}
\item[{\rm (1)}] $\ell(s) =\ell (s')$;
\item[{\rm (2)}] If
$\ok{s\sra t}$ 
then there exist $t_0,...,t_k\in \Sigma$, with $k\geq 0$, such that: (i)
$t_0=s'$; (ii)  for all $i\in [0,k-1]$, $\ok{t_i
\sra  t_{i+i}}$ and $s R t_i$; 
(iii) $t R t_k$;
\item[{\rm (3)}] $s'Rs$, i.e.\ $R$ is symmetric. 
\end{itemize}
Observe that condition~(2) allows the case $k=0$
and this simply boils down to requiring that $t R s'$.  
It turns out that the empty
relation is a dbs relation and dbs relations are closed under union. Hence,
the 
largest dbs  relation exists and 
is an equivalence relation called dbs equivalence, whose corresponding
partition is 
denoted by $P_{\mathrm{dbs}}\in \Part(\Sigma)$.

We showed in \cite{rt06} that $P_{\mathrm{dbs}}$
can be characterized as the coarsest strongly preserving partition  
$P_\fL$ for the following language $\fL$:
$$\varphi ::= 
~p ~|~ \varphi_1 \wedge \varphi_2 ~|~ \neg \varphi ~|~ 
\mathrm{EU}(\varphi_1,\varphi_2)$$
where the semantics $\beu:\wp(\Sigma)^2\ra
\wp(\Sigma)$ of the existential until operator $\mathrm{EU}$
is as usual:

\begin{tabbing} 
\indent
$\beu(S_1,S_2) = S_2 \cup \{s\in S_1 ~|~$\=$\exists s_0,...,s_n \in
\Sigma, \text{ with } n\geq 0, \text{ such that~(i)~} s_0=s,$\\
\>${\rm ~(ii)~} \forall i\in [0,n).\,
s_i\in S_1,\; s_i  \sra  s_{i+1},$
${\rm ~(iii)~} s_n\in S_2\}$.
\end{tabbing}

Therefore, as a straight instance of Corollary~\ref{partmain}, it turns out
that $\ok{\GPT^{\Part}_\beu (P_\ell)=P_{\fL}=
P_{\mathrm{dbs}}}$.

Groote and Vaandrager \cite{gv90} designed a  partition
refinement algorithm, here~denoted
by $\GV$, for
computing the partition $P_{\mathrm{dbs}}$. This algorithm uses the
following definitions of split and refiner:\footnote{In \cite{gv90},
$\pos(B_1,B_2)$ denotes $\beu(B_1,B_2)\cap B_1$.}
For any $P\in \Part(\Sigma)$ and $B_1,B_2\in P$,
$$
\begin{array}{lcl}
  \gvsplit(\tuple{B_1,B_2},P) & \ud & P\curlywedge \{\beu(B_1,B_2),
  \complement(\beu(B_1,B_2))\}\\
  \gvrefiners(P)&\ud &\{\tuple{B_1,B_2}\in 
  P\times P~|~\gvsplit(\tuple{B_1,B_2},P)\prec P\}.
\end{array}
$$

The algorithm $\GV$ is as follows. 
Groote and Vaandrager show how $\GV$ can be efficiently implemented in
$O(|\sra||\Sigma|)$-time.  
\begin{center}
{\small
$
\begin{array}{|l|}
\hline \\[-9pt]
~\mbox{{\bf input}}\!:~ \text{~partition~}P \in \Part(\Sigma);~~ \\
~\mbox{{\bf while~}} \gvrefiners(P)\neq \varnothing ~\mbox{{\bf do}}\\
~~~~~~\mbox{{\bf choose~}} \tuple{B_1,B_2} \in \gvrefiners(P);~~ \\
~~~~~~P:=\gvsplit(\tuple{B_1,B_2},P);\\[-1pt]
~\mbox{{\bf endwhile}}\\[-2pt]
~\mbox{{\bf output}}\!:~ P;
~~~~~~~~~~~~~~~~~~~~~~~
~~~~~~~~~~~~~~~~~~~~\framebox{$\GV$}\mbox{\hspace*{-5pt}}\\[-0.35pt]
\hline
\end{array}
$
}
\end{center}

It turns out that $\GV$ exactly coincides  with the optimized
instance 
$\IGPT_{\beu}^{\Part}$ that considers block refiners. 
This is obtained as a straight consequence of the following 
facts. 

\begin{lemma}\ \label{igv}
\begin{itemize}
\item[{\rm (1)}] $\gvrefiners(P) = \varnothing~$ iff $~\refiners_{\beu}^{\Part}(P)=
\varnothing$. 
\item[{\rm (2)}] $\gvsplit(\tuple{B_1,B_2},P) =
\refine_{\beu}^{\Part}(\tuple{B_1,B_2},P)$.
\end{itemize} 
\end{lemma}
\begin{proof}
{\rm (1)} It is sufficient to show that if for any $B_1,B_2\in P$,
  $\beu(B_1,B_2)\in\pad(P)$, then for any $S_1,S_2\in\pad(P)$,
  $\beu(S_1,S_2)\in\pad(P)$.
  Thus, we have to prove that for any
  $\{B_i\}_{i\in I},\{B_j\}_{j\in J} \subseteq P$, $\beu(\cup_i B_i,
  \cup_j B_j) = \cup_k B_k$, for some $\{B_k\}_{k\in K}\subseteq P$. 
  $\beu$ is an additive operator in its second argument, thus we only need to show
  that, for any $B\in P$, 
  $\beu(\cup_i B_i,B)=\cup_k B_k$, namely if $s\in \beu(\cup_i
  B_i,B)$ and $s\in B'$, for some $B'\in P$, then $B'\subseteq
  \beu(\cup_i B_i, B)$. 
  If $s\in\beu(\cup_i B_i, B)$, for some $B\in P$, then there exist
  $n\geq 0$ and   $s_0,...,s_n \in
  \Sigma$ such that $s_0=s$, $\forall
  j\in [0,n-1]. s_j\in \cup_i B_i$ and  $s_j  \sra
  s_{j+1}$, and   $s_n\in B$. 
  Let us prove by induction on $n$ that if $s'\in B'$ then 
  $s'\in\beu(\cup_i B_i, B)$.
  \begin{itemize}
  \item[--] $n=0$:
    In this case $s\in \cup_i B_i$ and $s\in B=B'$. Hence, for some
    $k$, $s\in B_k=B=B'$ and therefore $s\in \beu(B,B)=B$.
    Moreover, $\beu$ is monotone on its first argument and
    therefore $B'=B=\beu(B,B)\subseteq \beu(\cup_i B_i,B)$. 
    
  \item[--] $n+1$:
    Suppose that 
    there exist
    $s_0,...,s_{n+1} \in
    \Sigma$ such that $s_0=s$, $\forall
    j\in [0,n]. s_j\in \cup_i B_i$ and  $s_j  \ok{\sra}
    s_{j+1}$, and   $s_{n+1}\in B$. Let $s_n \in B_k$, for some $B_k\in
    \{B_i\}_{i\in I}$. Then, $s\in \beu(\cup_i B_i,B_k)$ and $s=s_0
    \ok{\sra} s_1 \ok{\sra} ... 
    \ok{\sra}
    s_n$. Since this trace has length $n$, 
    by inductive hypothesis, $s'\in \beu(\cup_i B_i,B_k)$. Hence,
    there exist $r_0,...,r_m \in \Sigma$, with $m\geq 0$, such that 
    $s'=r_0$, $\forall
    j\in [0,m-1]. r_j\in \cup_i B_i$ and  $r_j  \ok{\sra}
    r_{j+1}$, and   $r_{m}\in B_k$. Moreover, since $s_n
    \ok{\sra} s_{n+1}$, we have that 
    $s_n\in \beu(B_k,B)$. By
    hypothesis, $\beu(B_k,B)\supseteq B_k$, and therefore $r_m\in
    \beu(B_k,B)$. Thus, there exist $q_0,...,q_l \in \Sigma$, with $l\geq 0$, such that 
    $r_m=q_0$, $\forall
    j\in [0,l-1]. q_j\in B_k$ and  $q_j  \ok{\sra}
    q_{j+1}$, and   $q_{l}\in B$. We have thus find the following trace: 
    $s'=r_0 \ok{\sra} r_1 \ok{\sra}
    ... \ok{\sra} r_m = q_0 \ok{\sra} q_1
    \ok{\sra} ... \ok{\sra} q_l$, where all the
    states in the sequence but the last one $q_l$  belong to $\cup_i B_i$,
    while $q_l\in B$. This means that $s'\in \beu(\cup_i B_i,B)$.  
  \end{itemize}  

\noindent
{\rm (2)} 
By Point~(2) in Section~\ref{igptp},  
$\refine_{\beu}^{\Part}(\tuple{B_1,B_2},P)=
  P\curlywedge \{\beu(B_1,B_2),\complement (\beu(B_1,B_2))\} = 
\gvsplit(\tuple{B_1,B_2},P)$. 
\end{proof}

Hence, by Corollary~\ref{igpt}, we have that Lemma~\ref{igv}~(1)
allows us to exploit the $\IGPT_{\beu}^{\Part}$ algorithm in order to
choose refiners for $\beu$ among the pairs of blocks of the current
partition, so that by Lemma~\ref{igv}~(2) we obtain that
$\IGPT_{\beu}^{\Part}$ exactly coincides with the $\GV$ algorithm.

\subsection{A New Simulation Equivalence Algorithm}
It is well known that simulation equivalence is an appropriate state
equivalence to be used in abstract model checking because it strongly
preserves $\ACTLS$ and provides a better state-space reduction than
bisimulation equivalence.  However, computing simulation equivalence
is harder than bisimulation \cite{km02}.  A number of algorithms for
computing simulation equivalence exist, the most well known are  by Henzinger,
Henzinger and Kopke~\cite{hhk95}, Bloom and Paige~\cite{bp95}, Bustan
and Grumberg~\cite{bg03}, Tan and Cleaveland \cite{TC01} and
Gentilini, Piazza and Policriti~\cite{gpp03}.  The algorithms by
Henzinger, Henzinger and Kopke~\cite{hhk95} and Bloom and
Paige~\cite{bp95} run in $O(|\sra||\Sigma|)$-time and, as far as
time-complexity is concerned, they are the best available
algorithms. However, these algorithms have the drawback of a quadratic
space complexity that is limited from below by $O(|\Sigma|^2)$. The
algorithm by Gentilini, Piazza and Policriti~\cite{gpp03} appears to
be the best algorithm when both time and space complexities are taken
into account. Let $P_{\mathrm{sim}}$ denote the partition
corresponding to simulation equivalence so that
$|P_{\mathrm{sim}}|$ is the number of simulation 
equivalence classes. Then, Gentilini et al.'s algorithm runs in
$O(|P_{\mathrm{sim}}|^2 |\sra|)$-time while the space complexity is in
$O(|P_{\mathrm{sim}}|^2 + |\Sigma| \log (|P_{\mathrm{sim}}|))$. This
algorithm greatly improves Bustan and Grumberg's \cite{bg03} algorithm
in space while retaining the same time complexity. Moreover, Gentilini
et al.\ experimentally show that their algorithm also improves on Tan
and Cleaveland's~\cite{TC01} algorithm both in time and space while
the theoretical complexities cannot be easily compared. It is worth
remarking that all these algorithms are quite sofisticated and may use
complex data structures.  We show how $\GPT$ can be instantiated
in order to design a new simple and efficient simulation equivalence
algorithm with competitive 
space and time complexities  of, respectively, 
$O(|P_{\mathrm{sim}}|^2+|\Sigma|)$
and  $O(|P_{\mathrm{sim}}|^2 \cdot (|P_{\mathrm{sim}}|^2
+ |\sra|))$.

Consider a finite Kripke structure
$\cK=(\Sigma,\sra,\ell)$. 
A relation 
$R\subseteq \Sigma \times \Sigma $ is a
simulation on $\cK$ if for any $s,s'\in \Sigma $ such that $s R s'$: 
\begin{itemize}
\item[{\rm (1)}] $\ell(s') \subseteq \ell (s)$; 
\item[{\rm (2)}] For
any $t\in \Sigma $ such that $\ok{s \sra t}$, there exists $t'\in \Sigma $
such that $\ok{s'\sra t'}$
and $t R t'$.
\end{itemize}
Simulation equivalence 
$\sim_{\mathrm{sim}}\,\subseteq
\Sigma \times \Sigma $ is defined as follows: 
$s\sim_{\mathrm{sim}}s'$ iff there exist two simulation
relations $R_1$ and $R_2$ such that $s R_1 s'$ and $s' R_2s$. 
$P_{\mathrm{sim}}\in \Part(\Sigma)$ denotes
the partition corresponding to $\sim_{\mathrm{sim}}$.

It is known (see e.g.\
\cite[Section~8]{vg01}) that simulation equivalence on $\cK$ can be
characterized as the state equivalence induced by the following language $\fL$: 
$$\varphi ::= p ~|~ \varphi_1 \wedge \varphi_2 ~|~ \mathrm{EX}\varphi$$
namely, $P_{\mathrm{sim}}=P_\fL$, where the 
interpretation of $\mathrm{EX}$ in $\cK$ is the
standard predecessor operator.
Let us consider the GI $(\sd,
\Abs(\wp(\Sigma))_\sqsupseteq, \dAbs(\wp(\Sigma))_\sqsupseteq, \id)$
of disjunctive abstract domains into the lattice of abstract domains 
that we
defined in Example~\ref{estre}. 
As observed in
Example~\ref{estre},
it turns out that  $\sd \circ
\pre^{\scriptscriptstyle \cM}\circ\: \sd = \sd \circ
\pre^{\scriptscriptstyle \cM}$, namely the abstraction $\dAbs(\wp(\Sigma))$ is
backward complete for $\pre^{\scriptscriptstyle \cM}$. Thus, by applying
Theorem~\ref{main}~(i) we obtain 
$$\ok{\GPT_{\pre}^{\dAbs}(P_\ell)} = \ok{\sd
(\sS_{\pre}(\pad(P_\ell)))}.$$ 
In turn, by applying the partitioning 
abstraction $\pr$ we obtain $$\ok{\pr(\GPT_{\pre}^{\dAbs}
(P_\ell))}=\ok{\pr (\sd (\sS_{\pre}(\pad(P_\ell))))} = 
\ok{\pr (\sS_{\pre}(\pad(P_\ell)))}$$
because $\pr \circ \sd = \pr$.   
Also, by
(\ref{coro2}), we know that $\ok{\pr (\sS_{\pre}(\pad(P_\ell)))} = P_\fL=P_{\mathrm{sim}}$.
We have therefore shown that 
$$\ok{\pr(\GPT_{\pre}^{\dAbs}(P_\ell))}=P_{\mathrm{sim}}$$ 
namely the following instance 
$\ok{\GPT_{\pre}^{\dAbs}}$ allows to compute simulation equivalence. 

\begin{center}
{\small
$
\begin{array}{|l|}
\hline \\[-9pt]
~\mbox{{\bf input}}\!:~ \text{~disjunctive abstract domain~}
A := \sd(\{[s]_\ell\}_{s\in \Sigma})\in \dAbs(\wp(\Sigma));~ \\
~\mbox{{\bf while~}} (\refiners^{\dAbs}_{\pre} (A) \neq \varnothing) ~\mbox{{\bf do}}~\\
~~~~~~\mbox{{\bf choose~}} S \in \refiners^{\dAbs}_{\pre} (A );~ \\
~~~~~~A := \refine_{\pre}^{\dAbs} (S, A);\\
~\mbox{{\bf endwhile}}\\[-3.5pt]
~\mbox{{\bf output}}\!:~ A;
~~~~~~~~~~~~~~~~~~~~~~~~~~~~~~~~~~~~~~~~~~~~~~~~~~~~~~~~~~~~~~~~
~~~~~~~~~~~~~~~~~~~~~~~\framebox{$\GPT^{\dAbs}_{\pre}$}\mbox{\hspace*{-5.1pt}}\\[-0.35pt]
\hline
\end{array}
$
}
\end{center}

$\ok{\GPT^{\dAbs}_{\pre}}$
works by iteratively refining a disjunctive abstract domain
$A\in \dAbs(\wp(\Sigma))$, which is first initialized to the
disjunctive shell of the abstract domain determined by the labeling of
atoms. Then, $\ok{\GPT^{\dAbs}_{\pre}}$ iteratively finds a refiner $S$ for $A$,
namely a set $S\in \gamma(A)$ such
that   $\pre_\sra(S)$ does not belong to $\gamma(A)$ and therefore
may contribute to refine $A$, i.e.\ $\ok{\refine_{\pre}^{\dAbs} (S, A)}= 
\ok{\D(\gamma(A)\cup \pre_\sra(S))}
\sqsubset A$.
Simulation equivalence is then computed from the output
disjunctive abstract domain $A$ as $P_{\mathrm{sim}}=\pr(A)$.

It turns out that refiners of a disjunctive abstract domain $A$ 
can be chosen among images of blocks in $\pr(A)$,
namely in  
$$\subrefiners^{\dAbs}_{\pre} (A ) \ud \refiners^{\dAbs}_{\pre} (A )\cap
\{ \gamma(\alpha (B))~|~B\in \pr(A)\}.$$ 
In fact, since both $\gamma\circ \alpha$ and $\pre_\sra$ are
additive functions, it turns out that for any $S\in \gamma(A)$,
$\forall S \in \gamma(A).\: \pre_\sra(S)\in \gamma(A)$ iff $\forall
B\in \pr(A). \pre_\sra(\gamma(\alpha(B)))\in \gamma(A)$, so that
$\subrefiners^{\dAbs}_{\pre} (A ) = \varnothing$ iff  
$\refiners^{\dAbs}_{\pre} (A ) = \varnothing$, and therefore
Corollary~\ref{igpt} can be applied.

\subsubsection{A Data Structure for Disjunctive Abstract Domains}

It turns out that a disjunctive abstract domain $A_\leq 
\in \dAbs(\wp(\Sigma))$ can be represented through the
partition $\pr(A)\in \Part(\Sigma)$ induced by $A$ and 
the following relation 
$\unlhd_A$ on $\pr(A)$:
$$\forall B_1,B_2 \in \pr(A),~~ B_1 \unlhd_A B_2 \text{~~iff~~}
\gamma(\alpha(B_1)) \subseteq \gamma(\alpha(B_2)).$$

\noindent
It is clear that this gives rise to a partial order relation because if $B_1,B_2\in
\pr(A)$ and
$\gamma(\alpha(B_1))=\gamma(\alpha(B_{2}))$ then we can pick up
$s_1\in B_1$ and $s_2\in B_2$ 
so that $\gamma(\alpha(\{s_1\}))=\gamma(\alpha(B_1))=\gamma(\alpha(B_2))=
\gamma(\alpha(\{s_2\}))$, namely $s_1$ and $s_2$ are equivalent
according to $\pr(A)$ and therefore $B_{1}= B_{2}$. The
poset $\tuple{\pr(A),\unlhd_A}$ is denoted by $\poset(A)$. 
It turns out that a disjunctive abstract domain can always
be represented by this poset, namely
the closure operator induced by $A$ can be defined in
terms of $\poset(A)$ as follows.

\begin{lemma}
  \label{image}
  Let $A\in \dAbs(\wp(\Sigma))$. For any $S\subseteq\Sigma$, 
  $\gamma_A(\alpha_A(S))=\cup \{B\in\pr(A)~|~\exists C \in \pr(A).\, C \cap S \neq
\varnothing \;\:\&\;\: B\unlhd_A C\}$.
\end{lemma}
\begin{proof}
$(\subseteq)$ Consider any $x\in \gamma_A(\alpha_A (S))= \cup_{s\in
S}\gamma_A(\alpha_A(\{s\}))$. Then, there exists some $s\in S$ such
that $x\in \gamma_A(\alpha_A(\{s\}))$. We consider $B_x,B_s\in \pr(A)$
such that $x\in B_x$ and $s \in B_s$. Then, $B_s \cap S\neq
\varnothing$ and $B_x \unlhd_A B_s$ because $\gamma_A (\alpha_A (B_x))
= \gamma_A (\alpha_A (\{x\})) \subseteq \gamma_A (\alpha_A (\{s\})) =
\gamma_A (\alpha_A (B_s))$. \\
$(\supseteq)$ Let $B,C\in \pr(A)$ such that $s\in C\cap S$ and
$B \unlhd_A C$. Then, $B\subseteq \gamma_A (\alpha_A (B)) \subseteq 
\gamma_A (\alpha_A (C)) = \gamma_A (\alpha_A (\{s\})) \subseteq
\gamma_A (\alpha_A (S))$. 
\end{proof}

\begin{figure}[t]
\begin{center}
\begin{tabular}{|c|c|c|c|}
\hline
    \mbox{
      \xymatrix@R=10pt@C=10pt{
\\
        [123]\\
        [45]\ar@{-}[u]\\
\poset(A_1)
      }
    }
~~~&~~~
    \mbox{
      \xymatrix@R=13pt@C=1pt{
\mbox{~~}\\ 
\mbox{~~}\\
        [123] ~~~~ [45]\\
\poset(A_2)
      }
    }
~~~&~~~
    \mbox{
      \xymatrix@R=10pt@C=0pt{
&\mbox{~~}&\\
        &[123]&\\
        [4]\ar@{-}[ur] & & [5]\ar@{-}[ul]\\
&\poset(A_3)&
      }
    }
~~~&~~~
    \mbox{
      \xymatrix@R=9pt@C=0pt{
        &[3]&\\
        [1]\ar@{-}[ur] & & [2]\ar@{-}[ul]\\
&[45]\ar@{-}[ur]\ar@{-}[ul]&
\\
&\poset(A_4)&
      }
    }
\\ \hline
\end{tabular}
\end{center}
\caption{Disjunctive Abstract Domains as Posets.}
\label{figdue}
\end{figure}
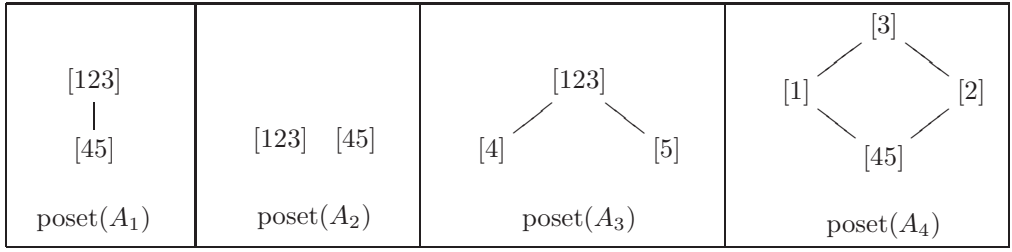

\begin{example}
  \label{exampleposet}
Some examples of posets that represent disjunctive abstract
domains are depicted in Figure~\ref{figdue}.
\begin{enumerate}
\item The disjunctive abstract domain $A_1 = \{\varnothing,
[45], [12345]\}$ is such that
$\pr(A_1)=\{[123],[45]\}$. 
\item The disjunctive domain $A_2
= \{\varnothing, [45], [123], [12345]\}$ induces the same
partition $\{[123],[45]\}$, while $\poset(A_2)$ is
discrete. 
\item  The disjunctive abstract domain $A_3 = \{\varnothing, [4],
[5], [45], [12345]\}$ induces the partition
$\pr(A_3)=\{[123],[4], [5]\}$.
\item The disjunctive abstract domain $A_4 = \{\varnothing,
[45], [145], [245], [1245], [12345] \}$ induces the partition
$\pr(A_4)=\{[1],[2],[3], [45]\}$. \qed
\end{enumerate}    
\end{example}

A disjunctive abstract domain $A\in \dAbs(\wp(\Sigma))$ 
is thus represented by $\poset(A)$. This means that 
our implementation of $\ok{\GPT^{\dAbs}_{\pre}}$ 
maintains and refines a partition $\pr(A)$ and an order relation on $\pr(A)$.
Let us describe how this can be done.

\subsubsection{Implementation}
Any state $s\in\Sigma$ is represented by a record \verb|State| that 
contains a pointer field \verb|block| that points to the block of the current
partition $\pr(A)$ that includes $s$ and a field \verb|pre| that represents
$\pre_\sra(\{s\})$ as a list of
pointers to the states in $\pre_\sra(\{s\})$.
The whole state space $\Sigma$ is represented as a doubly linked list \verb|states|
of \verb|State| so that insertion/removal can be done in $O(1)$. 
The ordering in the list \verb|states| matters
and may change during computation. 

Any block $B$
of the partition $\pr(A)\in \Part(\Sigma)$ is represented by a record 
\verb|Block| that contains the following fields: 
\begin{itemize}
\item[--]  \verb|first| and \verb|last| are pointers to \verb|State| such that
the block $B$ consists of all the states in the interval
[\verb|first|,\verb|last|] of the list
\verb|states|.
When a state is either added to
or removed from a block, the ordering in the list
\verb|states| changes accordingly and this can be done in $O(1)$. 
\item[--] \verb|less| is
a linked list of pointers to \verb|Block|. At the end of any
refinement step, the list \verb|less| for some block $B$
contains all the blocks $C\in \pr(A)$ which are less than or equal to $B$, i.e.\
such that $C\unlhd_A B$. In particular, the list \verb|less| is always nonempty
because \verb|less| always includes $B$ itself.
\item[--] \verb|intersection| is a pointer to
\verb|Block| which is set by the procedure
\verb|split| that  splits the current partition w.r.t.\ a set. 
\item[--] \verb|changedImage| is a boolean flag which is set by the procedure
\verb|orderUpdate|.
\end{itemize}

The blocks of the current partition $\pr(A)$ are represented as a doubly linked
list \verb|P| of \verb|Block|.

Let us face the problem of refining a disjunctive
abstract domain $A$ to $A'=\D(\gamma(A) \cup \{S\})$ for some $S\subseteq
\Sigma$.   If $P,P'\in \Part(\Sigma)$, $P'\preceq P$
and $B\in P'$ then let $\parent_P(B)\in P$ (when clear from the
context the subscript $P$ is omitted) denote the unique block
in $P$ (possibly $B$ itself)
that includes $B$. The following key result provides the basis for designing
an algorithm that updates
$\poset(A)$ to $\poset(A')$.

\begin{figure}[t]
\begin{center}
{\small
\begin{code}
\textbf[/* P is the current partition, S is a list of pointers to State */]
split(S) {
  \kforall state \kin S \kdo { 
     Block* B = state->block;
     \kif (B->intersection==NULL) \kthen {
        B->intersection = \knew Block;
        P.append(B->intersection);
        B->intersection->intersection = B->intersection; 
        B->intersection->less = copy(B->less);
        B->intersection->changedImage = false;
     }
     move(state,B,B->intersection);
     \kif (B = $\varnothing$) \kthen { \textbf[/* case: B $\subseteq$ S */]
        B->first = B->intersection->first; B->last = B->intersection->last;
        P.remove(B->intersection);
        \kdelete B->intersection
        B->intersection = B;
     }
  }
}
\end{code}
}

\vspace*{-5pt}
{\small
\begin{code}
\textbf[/* P is the current partition after a call to split(S) */]
orderUpdate() {
  \kforall B \kin P \kdo
     \kif (B$\cap$S = $\varnothing$) \kthen
        \kforall C \kin B->less 
           \kif (C $\neq \parent($C$)$) \kthen (B->less).append($\parent($C$)\cap$S);
     \kelse \textbf[/* case: B$\cap$S $\neq \varnothing$, i.e. B $\subseteq$ S */]
        \kforall C \kin B->less {
          \kif (C$\subseteq$S) \kthen \kcontinue;
          \textbf[/* case: C$\cap$S = $\varnothing$ */]
          (B->less).remove(C);
          \kif ($\parent($C$)\cap$S $\neq \varnothing$) (B->less).append($\parent($C$)\cap$S);
          B->changedImage = true;
        }
}
\end{code}
}
\end{center}
\caption{The procedures \texttt{split(S)} and \texttt{orderUpdate()}.}\label{figures}
\end{figure}

\begin{lemma}\label{refine}
Let $A \in \dAbs(\wp(\Sigma))$, $S\subseteq\Sigma$ and 
$A' = \D(\gamma(A) \cup \{S\})\in \dAbs(\wp(\Sigma))$. Let $P=\pr(A)\in
\Part(\Sigma)$ and  $P'=\ptsplit(S,P)\in \Part(\Sigma)$. 
Then, $\poset(A')=\tuple{P',\unlhd_{A'}}$, where for any $B',C'\in
P'$:
\begin{itemize} 
\item[{\text{\rm (i)}}]  
if $B'\cap S = \varnothing$ then $C'\unlhd_{A'} B' \:\Lra \: C'
\subseteq \gamma_A(\alpha_A(\parent(B')))$;
\item[{\text{\rm (ii)}}]  
if $B'\cap S \neq \varnothing$ then $C'\unlhd_{A'} B' \:\Lra \: C'
\subseteq \gamma_A(\alpha_A(\parent(B')))\cap S$.
\end{itemize}
\end{lemma}
\begin{proof}
Let $\mu = \gamma_A\circ \alpha_A$ and $\mu' = \gamma_{A'}\circ
\alpha_{A'}$. We first observe that if $x\in S$ then $\mu'(\{x\})
= \mu (\{x\})\cap S$, while if $x\not\in S$ then $\mu'(\{x\}) =
\mu(\{x\})$. 
We then show the following statement: for any $x,y\in \Sigma$, 
$$\mu'(\{x\}) \subseteq \mu'(\{y\}) \text{~~iff~~}
\mu(\{x\}) \subseteq \mu(\{y\}) \;\&\; (y\in S \,\Rightarrow x\in S)\eqno(*)$$

\noindent
($\Ra$) Since $\mu'\sqsubseteq \mu$, we have that $\mu \circ
\mu' = \mu$ so that $\mu(\{x\}) = \mu(\mu'(\{x\}))
\subseteq  \mu(\mu'(\{y\})) = \mu (\{y\})$. Moreover, if $y\in
S$ then $x\in \mu'(\{x\}) \subseteq \mu'(\{y\}) \subseteq
\mu'(S) = S$. 

\noindent
($\La$) If $y\in S$ then $x\in S$ so that $\mu' (\{x\}) =
\mu(\{x\}) \cap S \subseteq \mu(\{y\})\cap S = \mu' (\{y\})$. 
If instead $y\not\in S$ then $\mu'(\{x\}) \subseteq \mu(\{x\})
\subseteq \mu(\{y\}) = \mu'(\{y\})$. 

\medskip
\noindent
It is then simple to show that $P'= \ptsplit(S,P) =
\pr(A')$. In fact, $x \equiv_{A'} y$ iff $\mu'(\{x\}) =
\mu'(\{y\})$ and, by $(*)$, this happens iff $\mu(\{x\}) =
\mu(\{y\})$ and $x\in S \,\Lra\, y\in S$, namely iff $x$ and $y$
belong to the same block of $\ptsplit(S,P)$. 

\noindent
It is simple to derive from $(*)$ the following statement: for any
$B',C'\in P'$, 
$$\mu'(C') \subseteq \mu'(B')  \text{~~iff~~}
\mu(C') \subseteq \mu(B') \;\&\; (B'\cap S \neq \varnothing \,\Rightarrow\,
C' \cap S \neq \varnothing) \eqno(\ddagger)$$

\noindent
Let us now show points (i) and (ii). Let us observe that for any
$B'\in P'$, since $P'\preceq P=\pr(A)$, we have that $\mu(B')= \mu(\parent(B'))$.

\noindent
(i) Assume that $B'\cap S = \varnothing$. If $C' \unlhd_{A'} B'$,
i.e.\ $\mu'(C') \subseteq
\mu'(B')$, then, by $(\ddagger)$, $\mu(C') \subseteq \mu(B')$ so that
$C'\subseteq \mu(C') \subseteq \mu(B') = \mu(\parent(B'))$. On the
other hand, if $C' \subseteq \mu(\parent(B'))=\mu(B')$ then $\mu(C')
\subseteq \mu(B')$ and $B'\cap S \neq \varnothing \,\Ra\, C'\cap S
\neq \varnothing$ so that, by $(\ddagger)$, $\mu'(C') \subseteq
\mu'(B')$, i.e., $C' \unlhd_{A'} B'$. 

\noindent
(ii) Assume that $B'\cap S\neq \varnothing$. If $C' \unlhd_{A'} B'$,
i.e.\ $\mu'(C') \subseteq
\mu'(B')$, then, by $(\ddagger)$, $\mu(C') \subseteq \mu(B')$ and 
$C' \cap S \neq \varnothing$, namely $C' \subseteq S$. Also, $C'
\subseteq \mu(C')\subseteq \mu(B')=\mu(\parent(B'))$ so that $C'
\subseteq \mu(\parent(B'))\cap S$. On the other hand, if $C' \subseteq
\mu(\parent(B')) \cap S = \mu(B') \cap S$ then $C'\cap S\neq
\varnothing$. Also, from $C'\subseteq \mu(B')$ we obtain $\mu(C')
\subseteq \mu(B')$. Thus, by $(\ddagger)$, we obtain
$\mu'(C')\subseteq \mu'(B')$, i.e.\ $C' \unlhd_{A'} B'$. 
\end{proof}

A refinement step $\refine_{\pre}^{\dAbs} (S, A)=A'$ is thus implemented
through
the following two main steps:
\begin{itemize}
\item[{\rm (A)}] Update the partition  $\pr(A)$ to $\ptsplit(S,\pr(A))$;
\item[{\rm (B)}] Update the order relation $\unlhd_A$ on $\pr(A)$ to
$\unlhd_{A'}$ on $\ptsplit(S,\pr(A))$ using Lemma~\ref{refine}.
\end{itemize}

The procedure $\verb|split(|S\verb|)|$ in Figure~\ref{figures} 
splits the current partition $P\in \Part(\Sigma)$ w.r.t.\ a 
splitter $S\subseteq \Sigma$. Initially, each block $B\in P$ has the field
\verb|intersection| set to \verb|NULL|. At the end of
$\verb|split(|S\verb|)|$, the partition $P$ is updated to
$P'=\ptsplit(S,P)$ where for any $B\in P$:
\begin{itemize}
\item[--] If $\varnothing \subsetneq B\cap S \subsetneq B$ then $B$
is modified to $B\smallsetminus S$ by repeating the move statement in
line 12
and the newly allocated  block $B\cap S$ in line 6 is
appended in line 7 at the end of the current list of blocks;
\item[--] If $B\cap S=B$ or $B\cap S=\varnothing$ then $B$ is not
modified. 
\end{itemize}

\noindent
Moreover, the field
\verb|intersection| of
any $B'\in  P'=\ptsplit(S,P)$  is set as follows:

\begin{enumerate}
\item[(1)] If $B'\in P\cap P'$ and $B'\cap S=\varnothing$ then  
$B'\verb|->intersection|=\verb|NULL|$ because $\verb|split(|S\verb|)|$
does not modify the record $B'$. 
\item[(2)] If $B'\in P\cap P'$ and $B'\cap S\neq \varnothing$ (i.e.,
$B'\subseteq S$) then  
$B'\verb|->intersection|=B'$ (line 17).
\item[(3)] If $B'\in P'\smallsetminus P$ and $B'\cap S = \varnothing$ (i.e.,
$B' = \parent(B')\smallsetminus S$) then  
$B'\verb|->intersection|=\parent(B')\cap S$ (line 6).
\item[(4)] If $B'\in P'\smallsetminus P$ and $B'\cap S \neq \varnothing$ (i.e.,
$B' = \parent(B')\cap S$) then  
$B'\verb|->intersection|=B'$ (line 8).
\end{enumerate}

Note that for the ``old'' blocks in $P$,
$\verb|split(|S\verb|)|$ does not modify 
the corresponding list of pointers \verb|less|, while
the list \verb|less| for
a newly allocated block $B\cap S$ is a copy 
of the list \verb|less| of $B$ (line 9). Also observe that blocks that are
referenced by
pointers in some \verb|less| field may well be modified.

The procedure $\verb|orderUpdate()|$ in Figure~\ref{figures}
is called after \verb|split(S)| to update the \verb|less| fields in
order to represent the refined poset $\tuple{P',\unlhd_{A'}}$ defined in
Lemma~\ref{refine}. 
By exploiting the above points (1)-(4), 
let us observe the following points about the procedure
\verb|orderUpdate()| whose current partition represents $P'=\ptsplit(S,P)$.
\begin{enumerate}
\item[(5)] For all blocks $B'\in P'$, the test $B'\cap S= \varnothing$
in line 4
is translated as $B' \verb|->intersection| \neq B'$.
\item[(6)] The test $C \neq \parent(C)$ in line 6 is
translated as $C \verb|->intersection| \neq \verb|NULL|$ and\\ $C
\verb|->intersection| \neq C$.
\item[(7)] The block $\parent(C)\cap S$ in lines 6 and 10 is $C\verb|->intersection|$.  
\item[(8)] The test $C \subseteq S $ in line 9 is
equivalent to $C \cap S \neq \varnothing$ and is thus
translated as $C \verb|->intersection| = C$.
\item[(9)] Lines 4-6 implement the case (i) of Lemma~\ref{refine}. 
\item[(10)] Lines 7-14 implement the case (ii) of Lemma~\ref{refine}. 
\end{enumerate}

Moreover, if for some blocks $B,C\in P'$ we
have that $B\subseteq S$ and
$C$ belongs to the list $B\verb|->less|$ and $C\cap S =
\varnothing$~---~namely, we are in the case of line 10~---~then,
by Lemma~\ref{refine}, 
$\gamma_{A'}(\alpha_{A'}(B)) \subsetneq 
\gamma_A(\alpha_A(B))$, that is 
the image of $B$ changed. For these blocks $B$,
the flag $B\verb|->changedImage|$ is set to $\verb|true|$.

Finally, let us notice that the sequence of disjunctive abstract domains
computed by some run of $\ok{\GPT_{\pre}^{\dAbs}}$ is decreasing, 
namely  if $A$ and 
$A'$ are, respectively,
the current and next disjunctive abstract
domains then $A' \sqsubseteq 
A$. As a consequence, if 
an image $\gamma_{A}(\alpha_{A}(B))$, 
for some $B\in \pr(A)$, is not a
refiner for $A$ and $B$ remains a block in the next refined partition $\pr(A')$
then $\gamma_{A'}(\alpha_{A'}(B))$
cannot be a refiner for $A'$. 
Thus, a correct strategy for finding refiners consists in  
scanning the list of blocks of the current partition $P$ 
while in any refinement step   
from $A$ to $A'$, after calling
$\verb|split(S)|$,
all the blocks $B\in \pr(A')$ whose image
changed are 
moved to the tail of $P$. 
This leads to the implementation of $\ok{\GPT_{\pre}^{\dAbs}}$
described 
in Figure~\ref{figsim}. 

\begin{figure}[t]
{\small
\begin{code}
\textbf[/* the list Atoms represents the set] $\{\grasse[p]_\cK\subseteq \!\Sigma\;|\;p\in \AP\}$ \textbf[*/]
\textbf[/* P is initialized to the single block partition */]
Partition P = ($\Sigma$); $\Sigma$->less = {$\Sigma$};

\kforall S \kin Atoms \kdo {
   split(S); orderUpdate();
   split($\complement$S); orderUpdate();
}
\kforall B \kin P \kdo {  
   State* X = image(B); 
   State* S = NULL;
   \kforall s \kin X \kdo S.append(s->pre);
   split(S);
   orderUpdate();
   \kforall B \kin P \kdo {
      B->intersection = NULL;
      \kif (B->changedImage) {B->changedImage = false; P.moveAtTheEnd(B);}
   }
}
\end{code}  
\caption{Implementation of $\GPT_{\pre}^{\dAbs}$.}\label{figsim}
}
\end{figure}

\begin{theorem}\label{mainsim}
The algorithm in Figure~\ref{figsim}
 computes simulation equivalence
$P_{\mathrm{sim}}$ on $\cK$ in space
$O(|\Sigma|+|P_{\mathrm{sim}}|^2)$ 
and in time $O(|P_{\mathrm{sim}}|^2 \cdot (|P_{\mathrm{sim}}|^2
+ |\sra|))$. 
\end{theorem}
\begin{proof}
We have shown above that the algorithm in
Figure~\ref{figsim} is a correct implementation of 
$\ok{\GPT_{\pre}^{\dAbs}}$. Let us observe the following points.
\begin{itemize}
\item[{\rm (1)}] For any block $B\in P$, by Lemma~\ref{image},
$\verb|image(|B\verb|)|$ in line 11 can be computed in the worst case by
scanning each edge of the order relation $\unlhd_A$ on $P=\pr(A)$,  
namely in $O(|P|^2)$ time. Since any current partition is coarser than
$P_{\mathrm{sim}}$, it turns out that $\verb|image(|B\verb|)|$ can be
computed in $O(|P_{\mathrm{sim}}|^2)$ time.
\item[{\rm (2)}] The list of pointers $S$ in lines 12-13 representing
$\pre_{\sra}(\gamma_A(B))$ can be
computed in the worst case by traversing the whole 
transition relation, namely in 
$O(|\sra|)$-time.
\item[{\rm (3)}]
For any $S\subseteq \Sigma$,
$\verb|split(|S\verb|)|$ in line 14 is computed in 
$O(|S|)$ time.  
\item[{\rm (4)}]
$\verb|orderUpdate()|$ in line 15 is 
computed in the worst case by
scanning each edge of the order relation $\unlhd_A$ on $P=\pr(A)$,  
namely in $O(|P|^2)$ time, and therefore in 
$O(|P_{\mathrm{sim}}|^2)$ time.
\item[{\rm (5)}] The for loop in line 16 is computed in $O(|P|)$ time and 
therefore in 
$O(|P_{\mathrm{sim}}|)$ time.
\end{itemize}
Thus, an iteration of the for-loop takes
$O(2|P_{\mathrm{sim}}|^2 + |\sra| + |S| + |P_{\mathrm{sim}}|)$ time,
namely, because
$|S|\leq |\sra|$, $O(|P_{\mathrm{sim}}|^2 + |\sra|)$ time.

\noindent
In order to prove that the time complexity is 
$O(|P_{\mathrm{sim}}|^2 \cdot (|P_{\mathrm{sim}}|^2
+ |\sra|))$, let us show that 
the number of iterations of the for-loop is in
$O(|P_{\mathrm{sim}}|^2)$. Let $\{A_{i}\}_{i\in [1,k]} \in
\dAbs(\wp(\Sigma))$ be the sequence of different disjunctive abstract
domains computed in some run of the algorithm and let $\{\mu_i\}_{i\in
[1,k]} \uco(\wp(\Sigma))$ be the corresponding sequence of disjunctive
uco's. Thus, for any $i\in [1,k)$, $\mu_{i+1} \sqsubset \mu_{i}$ and
$P_{\mathrm{sim}}= \pr(\mu_k)$. Hence, for any $i\in [1,k]$,
$P_{\mathrm{sim}} \preceq \pr(\mu_i)$, so that for any $B\in
P_{\mathrm{sim}}$, $\mu_i(B)=\cup_{j\in J} B_j$ for some set of blocks
$\{B_j\}_{j\in J} \subseteq P_{\mathrm{sim}}$. We know that for any 
$i\in [1,k)$ there exists some block $B\in \pr(\mu_i)$ whose image
chages, namely $\mu_{i+1}(B) \subsetneq \mu_{i}(B)$. Note that 
$\mu_{i+1}(B) \subsetneq \mu_{i}(B)$ holds for some $B\in \pr(\mu_i)$
if and only if $\mu_{i+1} (B)
\subsetneq \mu_i (B)$ holds for some 
$B\in P_{\mathrm{sim}}$.
 Clearly, for any block $B\in P_{\mathrm{sim}}$,
this latter fact can happen at most $|P_{\mathrm{sim}}|$ times. Consequently, the
overall number of blocks that in some iteration of the for-loop change
image is bounded by 
$\sum_{B\in P_{\mathrm{sim}}} |P_{\mathrm{sim}}| =
|P_{\mathrm{sim}}|^2$. Hence, the overall number of blocks that are
scanned by the for-loop is bounded by $|\pr(\mu_1)| +
|P_{\mathrm{sim}}|^2$ and therefore 
the total number of iterations of the for-loop is in
$O(|P_{\mathrm{sim}}|^2)$.

\noindent
The input of the algorithm is the Kripke structure $\cK$, that is 
the list \verb|states| and for each state the list \verb|pre| of its
predecessors. In each iteration of the while loop we keep in memory
all the fields of the record \verb|State|, that need $O(|\Sigma|)$
space, 
the current partition, that needs $O(|P_{\mathrm{sim}}|)$ space, and the
order relation $\unlhd_A$, that needs $O(|P_{\mathrm{sim}}|^2)$
space. Thus,  the overall space complexity is
$O(|\Sigma|+|P_{\mathrm{sim}}|^2)$.  
\end{proof}

\subsection{A Language Expressing Reachability}
Let us consider the following language $\fL$ which is able to express
reachability together with propositional logic through the existential
``finally'' operator: 
$$\varphi::=p~|~\varphi_1 \wedge \varphi_2 ~|~\neg
\varphi~|~\mathrm{EF} \varphi$$ 
Given a Kripke structure $(\Sigma,\sra,\ell)$, 
the
interpretation $\bef:\wp(\Sigma)\ra \wp(\Sigma)$ of the reachability
operator $\mathrm{EF}$ is
as usual: $\bef(S)\ud \beu(\Sigma,S)$. 
Since $\fL$ includes propositional
logic, by   
Corollary~\ref{partmain}, it turns out that the instance $\ok{\GPT^{\Part}_{\bef}}$
allows to compute the coarsest strongly preserving partition $P_\fL$, namely 
$\ok{\GPT^{\Part}_{\bef}(P_\ell)} = P_\fL$.

It turns out that block refiners are enough, namely 
$$\ok{\blockrefiners_{\bef}^{\Part}
(P)}= \{B\in P~|~ P\curlywedge \{\bef(B),\ok{\complement(\bef(B))}\}\prec P
\}.$$ 
In fact, note that
$\ok{\blockrefiners_{ \bef}^{ \Part} (P)} = \varnothing$ iff
$\ok{\refiners_{ \bef}^{
\Part}(P)}=\varnothing$, so that,   
by exploiting Corollary~\ref{igpt}, we have that 
$\ok{\IGPT^{\Part}_{\bef}(P_\ell) = P_\fL}$. The optimized algorithm 
$\ok{\IGPT^{\Part}_{\bef}}$ is as follows. 
\begin{center}
{\small
$
\begin{array}{|l|}
\hline \\[-9pt]
~\mbox{{\bf input}}\!:~ \text{~partition~}P \in \Part(\Sigma);~ \\
~\mbox{{\bf while~}} (\blockrefiners^{\Part}_{\bef} (P) \neq \varnothing) ~\mbox{{\bf do}}~\\
~~~~~~\mbox{{\bf choose~}} 
B \in \blockrefiners^{\Part}_\bef (P);~ \\
~~~~~~P := P \curlywedge \{ \bef(B),\complement (\bef(B))\};\\[-1pt]
~\mbox{{\bf endwhile}}\\[-5pt]
~\mbox{{\bf output}}\!:~ P;
~~~~~~~~~
~~~~~~~~~~~~~~~~~~~~\framebox{$\IGPT^{\Part}_{\bef}$}\mbox{\hspace*{-5pt}}\\[-0.35pt]
\hline
\end{array}
$
}
\end{center}

\subsubsection{Implementation} The key point in implementing
$\ok{\IGPT^{\Part}_{\bef}}$ is 
the following property of ``stability under refinement'': 
for any $P,Q\in \Part(\Sigma)$, 
\begin{center}
if $Q \preceq P$
and $B \in P \cap Q$ 
then $P \!\curlywedge \{\bef(B),\!\ok{\complement(\bef(B))}\}\! =
P$ implies $Q\curlywedge \{\bef(B),\ok{\complement(\bef(B))}\} =
Q$.
\end{center}

\noindent 
As a consequence of this property, if some block $B$ of the current
partition $P_{\mathrm{curr}}$ is not a $\bef$-refiner for
$P_{\mathrm{curr}}$ and $B$ remains 
a block of the next partition $P_{\mathrm{next}}$ then $B$ cannot be a
$\bef$-refiner for $P_{\mathrm{next}}$.

\begin{figure}
\begin{center}
\begin{center}
{\small
$
\begin{array}{|l|}
\hline \\[-9pt]
~\mbox{{\bf input}}\!:\text{Transition System}~(\Sigma,\sra),\: 
                      \text{~List$\langle$Blocks$\rangle$~} P;\\
~(P_{\mathrm{scc}},\sra_{\mathrm{scc}}) := \text{scc}(\Sigma,\sra);\\
~\mbox{{\bf scan~}} B ~\mbox{\bf{in}}~ P ~\{\\
~~~~~~\text{List$\langle$BlocksOfBlocks$\rangle$}~B_{\mathrm{scc}}:=\{C\in P_{\mathrm{scc}}~|~B\cap C\not=\varnothing\};\\
~~~~~~\text{List$\langle$States$\rangle$}~S := \bigcup\mbox{\bf{computeEF}}(B_{\mathrm{scc}});\\
~~~~~~\mbox{\bf split}(S,P);\\
~\}\\[-3.5pt]
~\mbox{{\bf output}}\!: P;
~~~~~~~~~~~~~~~~~~~~~~~~~~~~~~~~~~~~~~~~~~~~~~~~~~~~~~~~~~~~~~~~\framebox{$\IGPT^{\Part}_{\bef}$}\mbox{\hspace*{-5.1pt}}\\[-0.35pt]
\hline
\end{array}
$
}
\end{center}

\bigskip
\begin{tabular}{cc}
\mbox{
{\small
$
\begin{array}{|l|}
\hline \\[-9pt]
~\text{List$\langle$States$\rangle$}~\mbox{{\bf compute}}\bef(\text{List$\langle$States$\rangle$}~S) \: \{ \\
~~~~~~\mbox{List$\langle$States$\rangle$}~\mathit{result};\\
~~~~~~\mbox{{\bf scan~}} s \mbox{{\bf~in~}} S ~\{\mathit{result}.{\text{append}}(s);~ \mbox{{\bf mark}}(s); \}\\
~~~~~~\mbox{{\bf scan~}} s \mbox{{\bf~in~}} \mathit{result} \\
~~~~~~~~~~~\mbox{{\bf forall}} ~r \in \pre(\{s\}) ~\mbox{{\bf do}} \\
~~~~~~~~~~~~~~~~\mbox{{\bf if}} ~(r ~\mbox{{\bf isNotMarked}})~ \mbox{{\bf then}}~ \{ \\
~~~~~~~~~~~~~~~~~~~~~\mathit{result}.\text{append}(r);~ \mbox{{\bf
    mark}}(r); \\
~~~~~~~~~~~~~~~~\}\\
~~~~~~\mbox{{\bf return}}~ \mathit{result};\\
~\}\\[1pt]
\hline
\end{array}
$
}
}
\hspace*{-5pt}&\hspace*{-5pt}
\mbox{
{\small
$
\begin{array}{|l|}
\hline \\[-9pt]
~\mbox{{\bf split}}(\text{List$\langle$States$\rangle$}~S,\,\text{List$\langle$Blocks$\rangle$}~P) \: \{ \\
~~~~~~\mbox{{\bf scan~}} s ~\mbox{\bf{in}}~ S ~\{\\
~~~~~~~~~~~\text{Block}~B := s.\text{block};\\
~~~~~~~~~~~\mbox{{\bf if~}} (B.\text{intersection} = \text{false})~\mbox{{\bf then}}~ \{\\
~~~~~~~~~~~~~~~~B.\text{intersection} := \text{true};~B.\text{split} := \text{true};\\
~~~~~~~~~~~~~~~~\text{Block}~ B\cap S := \mbox{\bf{new}}~ \text{Block};\\
~~~~~~~~~~~~~~~~P.\text{append}(B\cap S);\\
~~~~~~~~~~~\}\\
~~~~~~~~~~~\text{moveFromTo}(s,B,B\cap S);\\
~~~~~~~~~~~\mbox{{\bf if}} ~(B=\varnothing)~ \mbox{{\bf then}}~ \{ \\ 
~~~~~~~~~~~~~~~~B.\text{split}:=\text{false}; \\
~~~~~~~~~~~~~~~~B:=B\cap S; \\
~~~~~~~~~~~~~~~~P.\text{remove}(B\cap S);\\
~~~~~~~~~~~\}\\
~~~~~\}\\
~~~~~~\mbox{\bf{scan}}~ B ~\mbox{\bf{in}}~ P \\
~~~~~~~~~~~\mbox{\bf{if}}~(B.\text{split}=\text{true}) ~\mbox{\bf{then}}~ P.\text{moveAtTheEnd}(B);\\
~\}\\[1pt]
\hline
\end{array}
$
}
}
\end{tabular}
\end{center}
\caption{Implementation of $\IGPT^{\Part}_{\bef}$.}\label{fig4}
\end{figure}

This suggests an implementation of $\IGPT^{\Part}_{\bef}$ based on the following points: 
\begin{itemize}
\item[{\rm (1)}] The current partition $P$ is represented as a 
doubly linked list of blocks (so that a block removal can be done
in $O(1)$-time).
\item[{\rm (2)}] This list of blocks $P$ is scanned from the beginning 
 in order to find block refiners. 
\item[{\rm (3)}]  When a block $B$ of the current partition
$P$ is split into two new blocks $B_1$ and
$B_2$ then $B$ is removed from the list $P$ and $B_1$ and $B_2$
are appended at the end of $P$.
\end{itemize}
\noindent

These ideas lead to the implementation
$\IGPT^{\Part}_{\bef}$ described in Figure~\ref{fig4}.
As a preprocessing step we compute the DAG
of the strongly connected components (s.c.c.'s) of the directed graph
$(\Sigma,\sra)$, denoted by $(P_{\mathrm{scc}},\sra_{\mathrm{scc}})$.
This is done by the depth-first Tarjan's algorithm \cite{tar72} in
$O(|\sra|)$-time. 
This preprocessing step is done
because if $x\in \bef(S)$, for some $x\in \Sigma$ and 
$S\subseteq \Sigma$, then the whole block $B_x$ in the partition
$P_{\mathrm{scc}}$ that contains $x$~---~i.e., the 
strongly connected component containing
$x$~---~is contained in $\bef(S)$; moreover, let us also observe that 
$\bef(\{x\}) = \bef(B_x)$.  
The algorithm then proceeds by scanning the list of
blocks $P$ and performing the following three steps:
(1)~for the current block $B$ of the current partition $P$, 
we first compute the set $B_{\mathrm{scc}}$ 
of s.c.c.'s that contain some state in $B$; (2)~we then 
compute $\bef(B_{\mathrm{scc}})$ in the DAG
  $(P_{\mathrm{scc}},\sra_{\mathrm{scc}})$ because 
$\bef(B)=\bigcup \bef(B_{\mathrm{scc}})$; (3)~finally, 
we split the current partition $P$ w.r.t.\ the splitter $\bef(B)$.
The computation of $\bef(B_{\mathrm{scc}})$ 
is performed by the simple procedure $\mbox{\bf{computeEF}}(B_{\mathrm{scc}})$ in
Figure~\ref{fig4} in $O(|\sra_{\mathrm{scc}}|)$-time while 
splitting $P$ w.r.t.\ $S$ is done by the procedure  
$\mbox{\bf{split}}(S,P)$ in Figure~\ref{fig4} in $O(|S|)$-time.
It turns out that this implementation runs in $O(|\Sigma||\sra|)$-time.

\begin{theorem}
The implementation of $\IGPT^{\Part}_{\bef}$ in Figure~\ref{fig4} is
correct and runs
in $O(|\Sigma||\sra|)$-time. 
\end{theorem}
\begin{proof}
Let us show the following points.
\begin{itemize}
\item[{\rm (1)}]
Each iteration of the scan loop takes $O(|\sra|)$ time.
\item[{\rm (2)}]
The number of iterations of the scan loop is in $O(|\Sigma|)$.
\end{itemize}

\noindent
{\rm (1)} Let $B$ be the current block while scanning the current
partition $P$. The set $B_{\mathrm{scc}}=\{C\in
P_{\mathrm{scc}}~|~B\cap C \neq \varnothing\}$ is determined in
$O(|B|)$ time simply by scanning the states in $B$. The computation of 
$\bef(B_{\mathrm{scc}})$ in the DAG of s.c.c.'s
$(P_{\mathrm{scc}},\sra_{\mathrm{scc}})$ takes
$O(|\sra_{\mathrm{scc}}|)$ time, the union $S=\bigcup
\bef(B_{\mathrm{scc}})$ takes $O(|S|)$-time, while splitting $P$
w.r.t.\ $S$ takes $O(|S|)$ time. Thus, each iteration is done in 
$O(|B|+|\sra_{\mathrm{scc}}| +2|S|)=O(|\sra|+|\Sigma|)=O(|\sra|)$, since $|\Sigma|\leq
|\sra|$. 

\noindent
{\rm (2)}
Let $B$ be the current block of  
the current partition
$P_{\mathrm{curr}}$. Then, the next partition
$P_{\mathrm{next}}\preceq P_{\mathrm{curr}}$
is obtained by splitting through $\bef(B)$ a number $k\geq 0$ of
blocks of $P_{\mathrm{curr}}$ so that $|P_{\mathrm{next}}|=
|P_{\mathrm{curr}}|+k$, where we also consider the case that $\bef(B)$
is not a splitter for $P$, namely the case $k=0$.  
Recall that any partition $P$ has a certain 
height $\hbar(P)=|\Sigma|-|P|$ in the lattice
$\Part(\Sigma)$ which is bounded by $|\Sigma|-1$. Thus, after splitting $k$
blocks  we have that $\hbar(P_{\mathrm{next}}) = 
\hbar(P_{\mathrm{curr}}) - k$. The total number of blocks
which are split by some run of the algorithm is therefore bounded
by $|\Sigma|$. As a consequence, if $\{P_i\}_{i=0}^{m}$ is the sequence
of partitions computed by some run of the algorithm and $\{k_i\}_{i=0}^{m-1}$ is the 
corresponding sequence of
the number of
splits for each $P_i$, where $k_i\geq 0$,
then $\sum_{i=0}^{m-1} k_i  \leq |\Sigma|$. Also, at each
iteration $i$ the number of new blocks is $2 k_i$, so that the total
number of new blocks in some run of the algorithm is $\sum_{i=0}^{m-1} 2k_i \leq 2
|\Sigma|$. Summing up, the total number of blocks that are scanned by the scan
loop is $|P_0| + \sum_{i=0}^{m-1} 2k_i \leq |P_0| + 2|\Sigma| \leq 3
|\Sigma|$ and therefore the number of iterations is in $O(|\Sigma|)$.

\noindent
Since the computation of the DAG
of s.c.c.'s that precedes the scan loop takes 
$O(|\sra|)$-time, 
the overall time complexity of the algorithm is $O(|\sra||\Sigma|)$. 
\end{proof}

\begin{table}
  \centering
  \begin{tabular}{|l l l l l l l|}
    \hline
    Model&States&Transitions&Initial blocks&Final blocks&Blocks bisim.eq.&Time\\  
    \hline \hline
    cwi\_1\_2 & 4339 & 4774 & 27 & 27 &2959&0.05s\\
    \hline
    cwi\_3\_14 & 18548&29104&  3 &123 &123&1.29s\\
    \hline
    vasy\_0\_1 & 1513 &2448&  3 &12 &152&0.01s\\
    \hline
    vasy\_10\_56 & 67005&112312&13&18 &67005&0.89s\\
    \hline
    vasy\_1\_4 & 5647&8928&7&51&3372&0.16s\\
    \hline
    vasy\_18\_73 & 91789&146086&18&161&70209&8.98s\\
    \hline
    vasy\_25\_25 & 50433&50432&25217&50433&50433&721.37s\\
    \hline
    vasy\_40\_60 & 100013&120014&4&4&100013&0.69s\\
    \hline
    vasy\_5\_9 & 15162&19352&32&2528&13269&5.41s\\
    \hline
    vasy\_8\_24& 33290&48822&12&6295&30991&49.08s\\
    \hline
    vasy\_8\_38& 47345&76848&82&13246&47345&10.59s\\
    \hline
  \end{tabular} 
  \caption{Results of the experimental evaluation.}
  \label{results}
\end{table}

\subsubsection{Experimental Evaluation} 

A prototype of the above partition refinement algorithm 
$\IGPT^{\Part}_{\bef}$ has been developed in 
\mbox{C++}, whose source code is available at 
\mbox{http://www.math.unipd.it/$\sim$ranzato/GPT/IGPTPartEF.zip}. We
considered the well-known VLTS (Very Large Transition Systems) benchmark suite
for our experiments \cite{vasy}. 
The VLTS suite consists of transition systems encoded in the BCG
(Binary-Coded Graphs) format where labels are
attached to arcs.  Since our algorithm needs as input a Kripke
structure, namely a transition system where labels are attached to
states, we exploited a procedure designed by
Dovier et al.~\cite{dpp04} that transforms an edge-labelled graph $G$ into
a node-labelled graph $G'$ in a way such that bisimulation equivalences
on $G$  and $G'$ coincide. This conversion acts as follows:
any transition $\ok{s_1\xrightarrow{l}s_2}$ is replaced by two transitions 
$s_1\ra n$ and $n\ra s_2$, where $n$ is a new node labelled with $l$. 
Hence, this transformation grows the size of the graph: the number of
transitions is doubled and the number of nodes grows proportionally to
the average of the branching factor of $G$.

Our experimental evaluation of $\IGPT^{\Part}_{\bef}$ was carried out
on a Celeron 2.20 GHz laptop, with 512 MB RAM, running Linux 2.6.15
and GNU g++ 4.0.1. The results are summarised in
Table~\ref{results}, where we list the name of the original transition
system in the VLTS suite, the number of states and transitions of the
transformed transition system, the number of blocks of the initial
partition, the number of blocks of the final refined partition, the
number of bisimulation classes and the execution time of in
seconds. The experiments show that one can obtain significant 
state space reductions with a reasonable time cost. It can be
therefore interesting to experimentally evaluate whether this reduction can be 
practically  applied as a pre-processing step 
for checking reachability specifications.

\section{Related Work} 
Dams~\cite[Chapter~5]{dams96} presents a
generic splitting algorithm that, for a given language $\fL\subseteq
\ACTL$, 
computes an abstract model $A\in \Abs(\wp(\Sigma))$ that strongly
preserves $\fL$. This technique is inherently
different from ours, in particular because it is guided by a
splitting operation of an abstract state that depends on a given formula of
$\ACTL$. Additionally, Dams' methodology does not guarantee optimality of
the resulting strongly preserving abstract model, as instead we
do, because his algorithm may provide strongly preserving
models which are too concrete. Dams~\cite[Chapter~6]{dams96} also 
presents a generic partition refinement algorithm that
computes a given (behavioural) state equivalence and generalizes
$\PT$ (i.e., bisimulation equivalence) and Groote and Vaandrager
(i.e., stuttering equivalence) algorithms.      
This algorithm is parameterized on a notion of
splitter corresponding to some state equivalence, while our algorithm
is directly parameterized on a given language: the example 
given in \cite{dams96} (a ``flat'' version of $\CTLX$)  
seems to indicate that finding the right definition
of splitter for some language may be a hard task. Gentilini et
al.~\cite{gpp03} provide an algorithm that solves  a so-called generalized
coarsest partition problem, meaning that they generalized $\PT$
stability to partitions endowed with an acyclic relation 
(so-called partition pairs). 
They show that this technique can be instantiated to
obtain a logarithmic algorithm for $\PT$ stability and an efficient
algorithm for simulation equivalence. This approach is very
different from ours since the
partition refinement algorithm is not driven  by strong preservation
w.r.t.\ some language. 
Finally, it is also worth citing that Habib et al.~\cite{hpv99} show
that the technique of iteratively refining a partition by splitting
blocks w.r.t.\ some pivot set, as it is done in $\PT$, may be
generally applied for solving problems in various
contexts, ranging from strings to graphs. In fact, they show that a
generic skeleton of partition refinement algorithm, based on a
partition splitting step w.r.t.\ a generic pivot, can be instantiated
in a number of relevant cases where the context allows an appropriate
choice for the set of pivots.

\section{Conclusion and Future Work}
In model checking, the well known Paige-Tarjan algorithm is
used for minimally refining a given state partition in order to obtain
a standard abstract model that strongly preserves the branching-time
language $\CTL$ on some Kripke structure.  We designed a generalized
Paige-Tarjan algorithm, called $\GPT$, that minimally refines generic
abstract interpretation-based models in order to obtain strong
preservation for a generic inductive language. Abstract interpretation
has been the key tool for accomplishing this task.
$\GPT$ may be
systematically instantiated to classes of abstract models and
inductive languages that satisfy some conditions. We showed that
some existing partition refinement algorithms can be viewed as an instance
of $\GPT$ and that $\GPT$ may yield new efficient algorithms for
computing strongly preserving abstract models, like simulation
equivalence.  

$\GPT$ is parameteric on a domain of abstract models which is an
abstraction of the lattice of abstract domains $\Abs(\wp(\Sigma))$.
$\GPT$ has been instantiated to the lattice $\Part(\Sigma)$ of
partitions and to the lattice $\dAbs(\wp(\Sigma))$ of disjunctive
abstract domains. It is definitely interesting to investigate whether
the $\GPT$ scheme can be applied to new domains of abstract models. In
particular, models that are abstractions of $\Part(\Sigma)$ could be
useful for computing approximations of strongly preserving
partitions. As an example, if one is interested in reducing only a
portion $S \subseteq \Sigma$ of the state space $\Sigma$ then we may
consider the domain $\Part(S)$ of partitions of $S$ as an abstraction
of $\Part(\Sigma)$ in order to get strong preservation only on the
portion $S$.
 
\paragraph*{{\it Acknowledgments.}} 
This work was partially supported by the FIRB Project
``Abstract interpretation and model checking for the verification of
embedded systems'' and by the COFIN2004
Project ``AIDA: Abstract Interpretation Design and Applications''.
This paper is an extended and revised version of \cite{rt05}.


\begin{thebibliography}{99}



\bibitem{bp95}
B.~Bloom and R.~Paige. 
\newblock Transformational design and implementation of a new
efficient solution to the ready simulation problem.
\newblock \emph{Sci.\ Comp.\ Program.}, 24(3):189-220, 1995.


\bibitem{bcg88}
M.C.~Browne, E.M.\ Clarke and O.~Grumberg.
\newblock Characterizing finite {K}ripke structures in propositional
temporal logic.
\newblock \emph{Theor.\ Comp.\ Sci.}, 59:115-131, 1988.

\bibitem{bg03}
D.~Bustan and O.~Grumberg.
\newblock Simulation-based minimization.
\newblock \emph{ACM Trans.\ Comput.\ Log.},
4(2):181-204, 2003.


\bibitem{cgp99}
E.M.~Clarke, O.~Grumberg and D.A.~Peled.
\newblock \emph{Model Checking}.
\newblock The {M}{I}{T} Press, 1999.


\bibitem{CC77}
P.~Cousot and R.~Cousot.
\newblock Abstract interpretation: a unified lattice model for static analysis
  of programs by construction or approximation of fixpoints.
\newblock In \emph{Proc.\ 4$^{\mathit{th}}$ ACM POPL}, 238-252, 1977.

\bibitem{CC79}
P.~Cousot and R.~Cousot.
\newblock Systematic design of program analysis frameworks.
\newblock In \emph{Proc.\ 6$^{\mathit{th}}$ ACM POPL}, 269-282, 1979.

\bibitem{dams96}
D.~Dams. 
\newblock \emph{Abstract Interpretation and Partition Refinement for
Model Checking}. 
\newblock PhD Thesis, Eindhoven Univ., 1996.  

\bibitem{dnv95}
 R.~De Nicola and F.~Vaandrager.
\newblock Three logics for branching bisimulation.
\newblock \emph{J.\ ACM}, 42(2):458-487, 1995


\bibitem{dpp04}
A. Dovier, C.~Piazza and A.~Policriti.
\newblock An efficient algorithm for computing bisimulation equivalence.
\newblock \emph{Theor.\ Comput.\ Sci.}, 325(1):45-67, 2004.


\bibitem{fgr96}
G.~Fil\'e, R.~Giacobazzi and F.~Ranzato.
\newblock A unifying view of abstract domain design.
\newblock \emph{ACM Comput.\ Surv.}, 28(2):333-336, 1996.


\bibitem{gpp03}
R.~Gentilini, C.~Piazza and A.~Policriti.
\newblock From bisimulation to simulation: coarsest partition problems.
\newblock \emph{J.\ Automated Reasoning}, 31(1):73-103, 2003. 


\bibitem{gq01}
R.~Giacobazzi and E.~Quintarelli.
\newblock Incompleteness, counterexamples and refinements
in abstract model checking.
\newblock In \emph{Proc.\ 8$^{\mathit{th}}$ SAS}, LNCS~2126:356-373, 2001.

\bibitem{GR97}
R.~Giacobazzi and F.~Ranzato.
\newblock Refining and compressing abstract domains.
\newblock In \emph{Proc.\ 24th ICALP}, LNCS~1256, pp.~771-781,
Springer, 1997.


\bibitem{grs00}
R.~Giacobazzi, F.~Ranzato and F.~Scozzari.
\newblock Making abstract interpretations complete.
\newblock \emph{J.~ACM}, 47(2):361-416, 2000.

\bibitem{gv90}
J.F.~Groote and F.~Vaandrager.
\newblock An efficient algorithm for branching bisimulation and
stuttering equivalence.
\newblock In \emph{Proc.\ 17$^{\mathit{th}}$ ICALP}, LNCS~443:626-638, 1990.

\bibitem{hpv99}
M.~Habib, C.~Paul and L.~Vienot.
\newblock Partition refinement techniques: an interesting algorithmic
tool kit. 
\newblock \emph{Int.\ J.\ Found.\ Comput.\ Sci.}, 10(2):147-170, 1999.


\bibitem{hm85}
M.~Hennessy and R.~Milner.
\newblock Algebraic laws for nondeterminism and concurrency.
\newblock \emph{J.\ ACM}, 32(1):137-161, 1985.

\bibitem{hhk95}
M.R.~Henzinger, T.A.~Henzinger and P.W.~Kopke.
\newblock Computing simulations on finite and infinite graphs.
\newblock In \emph{Proc.\ 36$^{\mathit{th}}$ FOCS}, 453-462, 1995.

\bibitem{hmr05}
T.A.~Henzinger, R.~Maujumdar and J.-F.~Raskin.
\newblock A classification of symbolic transition systems.
\emph{ACM Trans.\ Comput.\ Log.}, 6(1):1-31, 2005.


\bibitem{km02}
A.~Kucera and R.~Mayr.
\newblock Why is simulation harder than bisimulation? 
\newblock In \emph{Proc.\ 13$^{\mathit{th}}$ CONCUR}, LNCS~2421:594-610, 2002.

\bibitem{lam83}
L.~Lamport. 
\newblock What good is temporal logic?
\newblock In \emph{Information Processing '83}, pp.~657-668,
IFIP North-Holland, 1983.


\bibitem{pt87}
  R.~Paige and R.E.~Tarjan.
  \newblock Three partition refinement algorithms.
  \newblock \emph{SIAM J.\ Comput.}, 16(6):973-989, 1987

\bibitem{rt05}
F.~Ranzato and F.~Tapparo.
\newblock An abstract interpretation-based refinement algorithm for
strong preservation. 
\newblock In \emph{Proc.\ 11th Intern.\ Conf.\ on Tools and Algorithms
for the Construction and Analysis of Systems (TACAS'05)}, LNCS~3440, pp.~140-156,
Springer, 2005.


\bibitem{rt06}
F.~Ranzato and F.~Tapparo.
\newblock Generalized strong preservation by abstract interpretation.
\newblock \emph{J.\ Logic and Computation}, to appear, 2006. Extended
abstract appeared in Proc.\ ESOP'04, LNCS~2986:18-32, 2004.

\bibitem{tar72}
R.E.~Tarjan. 
\newblock Depth-first search and linear graph algorithms.
\newblock In \emph{SIAM J.\ Comput.}, 1(2):146-160, 1972.

\bibitem{TC01}
L.~Tan and R.~Cleaveland.
\newblock Simulation revisited.
\newblock In In \emph{Proc.\ 7th Intern.\ Conf.\ on Tools and Algorithms
for the Construction and Analysis of Systems (TACAS'01)}, LNCS~2031,
pp.~480-495,
Springer, 2001.


\bibitem{vg01}
R.J.~van Glabbeek.
\newblock The linear time - branching time spectrum I: the semantics
of concrete sequential processes. 
\newblock In \emph{Handbook of Process Algebra}, pp.~3-99, Elsevier, 2001. 

\bibitem{vasy}
\newblock The VLTS Benchmark Suite.
\newblock \mbox{http://www.inrialpes.fr/vasy/cadp/resources/benchmark\_bcg.html}


\end{thebibliography}
\end{document}